\documentclass[journal]{IEEEtran}

\usepackage{cite}
\usepackage{multicol,lipsum}
\usepackage{amsmath,amssymb,amsfonts}
\usepackage{algorithmic}
\usepackage{graphicx}
\usepackage{textcomp}
\usepackage{xcolor}
\usepackage{booktabs}
 \usepackage{hyperref}
\usepackage{enumerate}
\usepackage{algorithm}
\usepackage{multirow}
\usepackage{standalone}
\usepackage{tikz}
\usetikzlibrary{positioning}
\usetikzlibrary{decorations.markings}

  \usepackage{enumerate}
\usepackage{subcaption}
 \usepackage{setspace}
\usepackage{verbatim}
   \usepackage{tabularx}
 \usepackage{makecell}

 \usepackage{hyperref}

\ifCLASSINFOpdf
\else
\fi
\begin{document}
%
\title{A Design of Low-Projection SCMA Codebooks  for Ultra-Low Decoding Complexity in Downlink IoT Networks }
%
%
%

\author{Qu Luo,   ~\IEEEmembership{Graduate Student Member,~IEEE,}
      Zilong Liu, ~\IEEEmembership{Senior Member,~IEEE,}
       Gaojie Chen, ~\IEEEmembership{Senior Member,~IEEE,} 
       Pei Xiao, ~\IEEEmembership{Senior Member,~IEEE,}
          Yi Ma, ~\IEEEmembership{Senior Member,~IEEE }
          and  Amine  Maaref,~\IEEEmembership{Senior Member,~IEEE.}  
\thanks{ Qu  Luo, Gaojie Chen,  Pei  Xiao  and Yi Ma are  with  5G \& 6G  Innovation Centre, Institute for Communication Systems (ICS), University of Surrey, UK, email:\{q.u.luo,  gaojie.chen, p.xiao, m.yi\}@surrey.ac.uk.}
\thanks{ Zilong   Liu   is   with   the   School   of   Computer   Science   and   Electronic   Engineering,   University   of   Essex,   UK,    email:   zilong.liu@essex.ac.uk.}
\thanks{ Amine  Maaref  is   with   the  Canada Research Center, Huawei Technologies Company Ltd., Ottawa, Canada,  email:  amine.maaref@huawei.com.}
 }

\markboth{ } 
{Qu \MakeLowercase{\textit{et al.}}: A  Design of Low-Projection  SC Codebooks   that Enables Low Decoding Complexity for Downlink SCMA}%
%

\maketitle

\begin{abstract} This paper conceives a novel sparse code multiple access (SCMA) codebook design which is motivated by the strong need for providing ultra-low decoding complexity and good error performance in downlink Internet-of-things (IoT) networks, in which a massive number of low-end and low-cost IoT communication devices are served.    By focusing on the typical Rician fading channels, we analyze the pair-wise probability of superimposed SCMA codewords and then deduce the  design metrics for multi-dimensional constellation construction and  sparse codebook optimization.  For significant reduction of  the decoding complexity, we advocate the key idea of projecting the multi-dimensional constellation elements to a few overlapped complex numbers in each dimension, called low projection (LP). An emerging modulation scheme, called golden angle modulation (GAM), is considered for multi-stage LP optimization, where the resultant multi-dimensional constellation is called LP-GAM.   Our analysis and simulation results show the superiority of the proposed LP codebooks (LPCBs) including one-shot decoding convergence and excellent  error rate performance. 
In particular,  the proposed LPCBs lead to decoding complexity reduction by at least $97\%$ compared to that of the conventional codebooks,  whilst owning  large minimum Euclidean distance.  Some examples of the proposed  LPCBs are available at  \url{https://github.com/ethanlq/SCMA-codebook}. 
 

\end{abstract}


%
\IEEEpeerreviewmaketitle

\section{Introduction}

  \IEEEPARstart{T}{he}     Internet-of-things (IoT) represents a revolutionary paradigm shift from the legacy human-centric networks (e.g., in 3G and 4G) to massive machine-type communications, where the latter is a major use case in the 5G-and-beyond mobile networks  \cite{FuqahaCommun}.
  Under this big picture, however, 
it is challenging to support the concurrent  communications of  massive IoT devices.  Due to the limited time-frequency resources, traditional  orthogonal multiple access (OMA) may be infeasible.   A disruptive technique for addressing such a challenge  is called non-orthogonal multiple access (NOMA) which permits several times of IoT devices larger than that in OMA systems communicating simultaneously \cite{SparseYu}.

Existing NOMA techniques can be mainly categorized into two classes: power-domain NOMA \cite{islam2016power}  and code-domain NOMA (CD-NOMA)  \cite{liu2021sparse}. The former works by superimposing multiple users with distinctive  power levels over the  identical time-frequency resources, whereas the latter carries out multiplexing  by employing different  codebooks/sequences with certain intrinsic structural properties. 
This paper focuses on  the sparse code multiple access (SCMA) \cite{taherzadeh2014scma, hoshyar2008novel }, which is a  representative CD-NOMA scheme with sparse codebooks. In SCMA, every user's  instantaneous input message bits  are directly mapped to a multi-dimensional sparse codeword drawn   from a pre-designed codebook \cite{luo2021error}.  
 SCMA is attractive as judiciously designed codebooks lead to constellation gain (due to signal space diversity), whilst their sparsity can be leveraged to enable near-optimum  message passing (MPA) decoding \cite{hoshyar2008novel}.
 
  Many studies have been carried out  to reap the benefits of SCMA for the enabling of massive connectivity in  IoT networks  \cite{AchievingZeng,SparseMoon,JointMiuccio,AnalyzingLai,ZhangComplexity,WangIterative}.  Recently, SCMA has also been applied in satellite IoT networks with novel  random access design \cite{ZhangUser,li2019asynchronous} and new receiver schemes  \cite{ZhangComplexity,WangIterative}. To realize these SCMA based IoT systems, nevertheless, a fundamental question is: \textit{how to design new sparse codebooks to enable highly efficient decoding in terms of complexity, energy/storage consumption, and convergence?} We will address this as detailed in the sequel.

 \subsection{Related works}




  In SCMA literature, to avoid the excessive design complexity, a widely adopted approach is to apply certain  user-specific operations (e.g., interleaving,    permutation, shuffling and phase rotations) to a common multidimensional constellation, called a   mother constellation (MC), for multiple sparse codebooks   \cite{nikopour2013sparse,taherzadeh2014scma}.   Following this approach, various types of SCMA codebooks have been developed in \cite{mheich2018design,cai2016multi,deka2020design,yu2015optimized,Zhang,li2020design,chen2020design}.  For excellent error performance, it is desirable to maximize the   minimum product distance (MPD) or minimum Euclidean distance (MED) of an MC \cite{boutros1996good}. In general, a large MED  leads to reliable detection in the Gaussian   channel, whereas a large MPD is preferred for robust transmissions in the Rayleigh fading channel \cite{WenDesigning}.
For example, the authors in \cite{cai2016multi} proposed a simple constellation rotation and interleaving  codebook design scheme  by using  an MC with enlarged MED.  
Golden angle modulation (GAM)  constellation was adopted in   \cite{mheich2018design}  to construct SCMA codebooks with low  peak-to-average power ratio (PAPR) properties.  
In \cite{yu2015optimized}, Star-QAM constellation was introduced to construct MC with large MED for downlink SCMA.   
In addition, a  uniquely decomposable constellation group   based codebook design approach was developed  in \cite{Zhang} by maximizing the MED at each resource node. With the aid of generic algorithm, in \cite{li2020design}, a  novel class of power-imbalanced SCMA codebooks  was developed in \cite{li2020design} by maximizing    the MED of the superimposed codewords whilst maintaining a large  MPD of the MC.   Recently, new downlink  quaternary sparse codebooks with large MED were obtained in \cite{huang2021downlink}  by a new iterative algorithm based on alternating maximization with exact penalty.     

It is worth mentioning that  the complexity  of MPA decoding can be significantly  reduced by  designing an MC  with low projection   (LP) numbers \cite{bayesteh2015low,bao2017bit}. This is achieved by allowing  certain overlapped constellation elements at the same dimension. The rationale is that, due to the multidimensional nature, any two MC vectors with certain overlapping dimension(s) may be well separated in other  dimension(s). Importantly, constellation overlapping at certain dimension (or more) leads to less computations in the message passing based inference.  By leveraging this advantage, the authors in \cite{bayesteh2015low}  proposed a modified LP-MPA decoder which  can utilise the  constellation structure to reduce    the  number of calculations of  belief messages  at each iteration.     Subsequently, the authors in   \cite{bao2017bit} presented an LP codebook design based on   the amplitude phase-shift keying   constellation in each dimension in uplink SCMA systems.  
To the best of our knowledge, a comparative study on sparse codebooks  with LP numbers in downlink channels is still missing.  This is challenging because one  needs to deal with the difficulties for joint optimization of the MC and user-specific codebook operators as well as the prohibitively high computational complexity in calculating the minimum Euclidean/product distance  of  the superimposed codewords.

 
 \subsection{Motivations and contributions} 

This paper is driven by the key observation that a majority of IoT communication devices  are usually low-end and lost-cost wireless sensors which are very constrained in terms of resources such as energy, CPU, and memory/storage capacity\footnote{One may refer to  \cite{HahmOperating} for an excellent survey on low-end IoT devices as well as their major challenges posed to the operating system design.}.
In view of this and unlike many existing SCMA papers focusing on efficient receiver design, we aim for achieving extremely low decoding complexity and super-rapid convergence rate at the receiver from the codebook design perspective. In addition, in spite of numerous SCMA codebook designs for Gaussian channels or Rayleigh fading channels, they may not be optimal for other real-life  channels.   Typically, a practical wireless channel consists of a  small-scale fading caused by non-line of sight (NLoS) propagation and static line of sight (LoS) path, which can be modeled as Rician fading channels.  The Rician fading channels are widely present  in practical IoT networks, such as the  terrestrial IoT networks \cite{ZhangPerformance,YuCost},   satellite IoT  networks \cite{5G_NR_s_iot,lutz1991land, vucetic1992channel,li2019asynchronous,gui2021scma,ZhangComplexity,WangIterative}, and unmanned aerial vehicle (UAV)  networks \cite{ErnestNOMA,WangTrajectory}. 

The above observations     motivate  us to design   enhanced codebooks  that  can substantially  reduce the decoding complexity, while   achieving good error   performance for SCMA based downlink IoT networks under Rician  fading channels. The main contributions of this work are summarized as follows:

\begin{itemize}
  \item 
  We analyze  the  system performance of SCMA in downlink Rician fading channels, and  derive the minimum distance of the  superimposed  codewords   by investigating  and minimizing the conditional  PEP. It is shown that our   derived minimum distance  generalizes  the previous design metrics, i.e., the MED and MPD, which are for the AWGN and Rayleigh fading channels, respectively.

  \item We propose a novel class of LP codebook (LPCB) to   enable  ultra-low   decoding complexity  and   excellent   error
performance.  Specifically, an efficient algorithm  is first proposed to  construct a novel  one-dimensional basic constellation with  proper LP numbers based on GAM  (LP-GAM)\footnote{GAM is an emerging  modulation that  leads to higher  mutual information and lower  PAPR performance over pulse-amplitude modulation   and square quadrature amplitude modulation (QAM) \cite{larsson2017golden}.}.  Afterwards, general multi-stage  scheme is developed to construct multiple sparse codebooks by jointly optimizing the MC, constellation operators, and bit labeling.


\item  We pay a special attention on the application of the proposed LPCBs to a challenging design case where the codebook size and overloading factor are both large. The conventional state-of-the-art, however, may not be effective due to the prohibitively high decoding complexity in this case.   Extensive simulations are   conducted to evaluate the proposed LPCBs. Interestingly, it is shown that a converged   decoding can be achieved in a single shot (called 
`one-shot decoding' in this paper) for some of the proposed LPCBs, leading to dramatic reduction of the decoding complexity.  In addition, to the best of our    knowledge, some of the proposed LPCBs also achieve the largest MED values among all existing codebooks.
\end{itemize}

 \subsection{Organization}

The rest of the paper is organized as follows. In Section II, the system model of downlink SCMA based IoT network along with  the  multiuser detection technique are presented.   Section III analyzes the PEP of SCMA based IoT system  under Rician fading channels and derives  the criteria of  sparse codebooks. In Section IV, the proposed sparse codebook  design and optimization are elaborated.  The numerical results are given in Section V. Finally, conclusions are made in Section VI.

 \subsection{Notations}

  $\mathbb{C}^{k\times n}$ and $\mathbb{B}^{k\times n}$ denote the $(k\times n)$-dimensional complex and binary matrix spaces, respectively.  ${{Z}_{{{M} }}}$ stands for the integer set $\left\{ 1,2,\cdots ,{{M} } \right\}$. 
$\lfloor \cdot \rfloor$, ${\mathrm{|}}\cdot {\mathrm{|}}$ and ${\mathrm{||}}\cdot {\mathrm{||}}$ denote the floor function, absolute value, and the $\mathcal   \ell _{2}-\mathrm {norm}$, respectively.  
$ \mathcal{CN}\left (0, 1\right )$  denotes complex Gaussian distribution with  zero-mean and unit-variance. diag$ (\cdot ) $ denotes the diagonalization of a matrix.

\section{System model}
 \label{sec2}
\subsection{Signal Model}

\begin{figure}[t]   
\centering 
\includegraphics[width=0.45\textwidth]{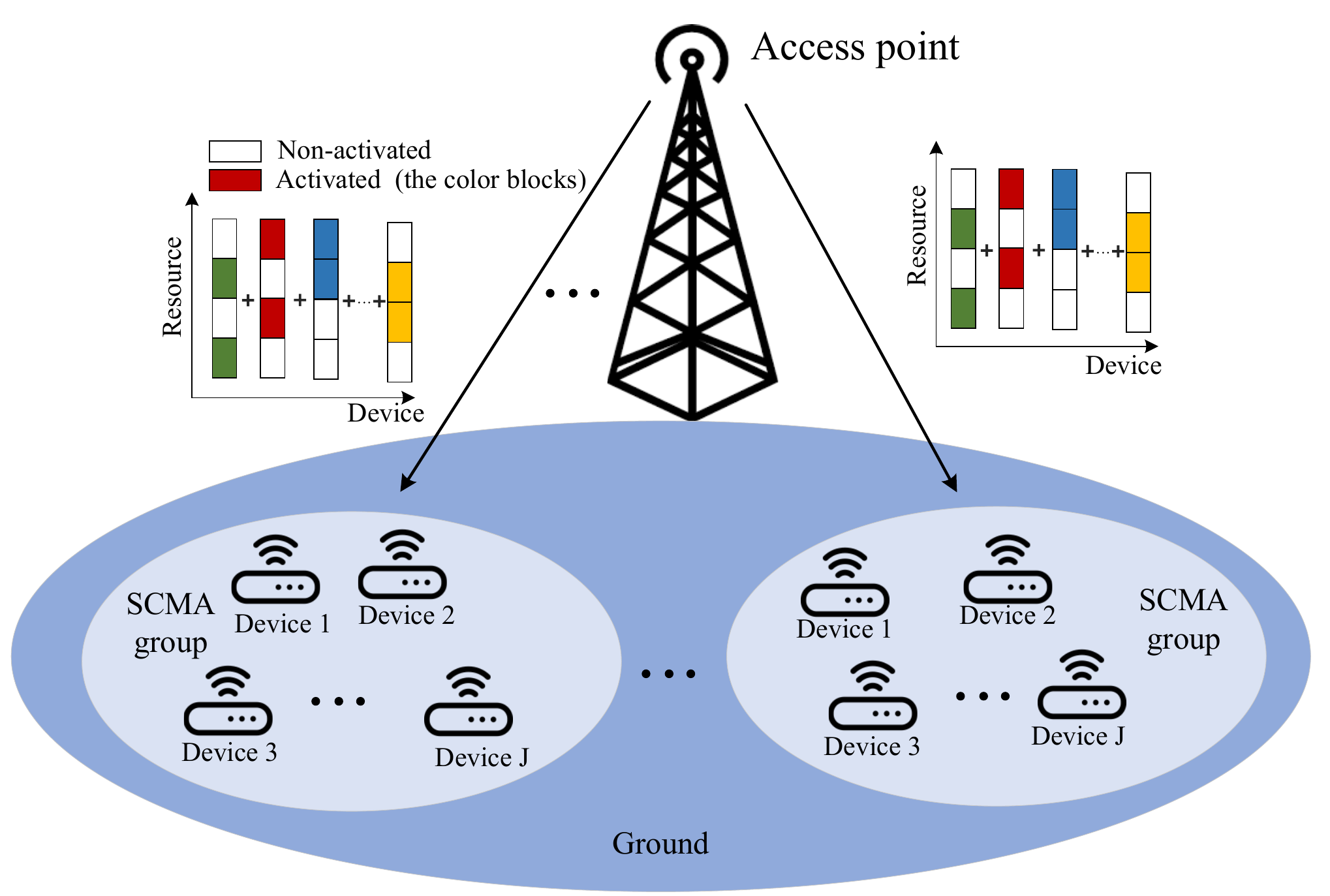} 
\caption{  An illustration of SCMA deployment   in an  IoT network.   } 
\label{FigSys}
 \end{figure}

 We consider an  SCMA system in  a  downlink IoT   network,  where the  access point  serves multiple SCMA   groups\footnote{ The users in the same group generally have similar locations, and the detailed user grouping  scheme is beyond the scope of this paper.}, as shown in Fig. \ref{FigSys}.    Each   SCMA   group consists of  $J$ IoT    devices  occupying $K$ orthogonal  resource nodes, and the overloading factor is defined   as $ \lambda = \frac{J}{K}> 100\% $.   At  the satellite side,  the SCMA encoder maps $\log_2\left(M\right)$   binary bits to a length-$K$ codeword $ \mathbf {x} _{j}$ drawn from pre-defined codebook $ \boldsymbol {\mathcal {X}}_{j}  \in \mathbb {C}^{K \times M}$, where $M$ denotes  the modulation order.  The mapping process is defined as  
\begin{equation} \label{scmamap}
 \small
 f_j:\mathbb{B}^{\log_2M \times 1}  \rightarrow { \boldsymbol {\mathcal X}}_{j}   \in \mathbb {C}^{K \times M}, {~\text {i.e., }}\mathbf {x}_{j}=f_{j}(\mathbf {b}_{j}), 
\end{equation}
  where $ \boldsymbol {\mathcal {X}}_{j}=\{\mathbf{x}_{j,1}, \mathbf{x}_{j,2},\ldots,\mathbf{x}_{j,M}\} $  is the codebook set for the $j$th user with cardinality of $M$ and  $\mathbf {b}_{j}=[b_{j,1},b_{j,2},\ldots,b_{j,\log _{2} M}]^{\text{T}}\in \mathbb{B}^{\log _{2} M \times 1}  $ stands for the $j$th user's incoming    binary message vector. 
The $K$-dimensional complex codewords in the SCMA codebook are sparse vectors with $N$ non-zero elements and $N < K$. 
Let $\mathbf {c}_{j}$ be a length-$N$ vector drawn from  $ \boldsymbol{ {\mathcal C}}_{j}\subset \mathbb {C}^{N \times M }$, where $   \boldsymbol{{\mathcal C}}_{j}$ is obtained  by removing all the zero elements  in  $ \boldsymbol{{ \mathcal X}}_{j}$.  We further define the mapping from $\mathbb{B}^{\log_2M}$ to  $ \boldsymbol{{\mathcal C}}_{j}$ as \cite{luo2022novel}
\begin{equation} \label{SCMAmapping}
 \small
g_{j}:\mathbb {B}^{\log _{2}M\times 1}\mapsto  \boldsymbol{{\mathcal C}}_{j}, \quad {~\text {i.e., }}\mathbf {c}_{j}=g_{j}(\mathbf {b}_{j}).
\end{equation}

Thus, the SCMA mapping defined in (\ref{scmamap}) now can be re-written as 
\begin{equation} 
 \small
\label{scmaMapping}
f_{j}:\equiv \mathbf {V}_{j}g_{j}, \quad {~\text {i.e., }}\mathbf {x}_{j}=\mathbf {V}_{j}g_{j}(\mathbf {b}_{j}),
\end{equation}
where $\mathbf {V}_{j} \in \mathbb {B}^{K \times N} $ is a  mapping   matrix that maps the $N$-dimensional vector  to a $K$-dimensional sparse SCMA codeword. The sparse structure of the $J$ SCMA codebooks can be represented by an  indicator (sparse) matrix  $\mathbf {F}_{K \times J} = \left [ \mathbf {f}_1, \ldots, \mathbf {f}_J \right] \subset \mathbb {B}^{K\times J}$ where  $\mathbf {f}_j = \text {diag}(\mathbf {V}_j\mathbf {V}_j^{\text{T}})$.  The 
variable node  $j$ is connected to   resource node  $k$ if and only if $f_{k,j}=1$.   Furthermore, let  $ \boldsymbol{\boldsymbol{\mathcal {I}}}_r(k)=
\left\{ {j\left| {{f_{k,j}} = 1} \right.} \right\}$ and $\boldsymbol{\boldsymbol{\mathcal {I}}}_u(j)=\left\{ {k\left| {{f_{k,j}} =1} \right.} \right\}$ are  the set of user indices sharing resource node $k$ and the set of resource indices occupied  by user $j$, respectively.  
In this paper, the following two factor indicator matrices with $\lambda = 150\%$ and $\lambda = 200\%$ are employed \cite{mheich2018design,li2020design}:
\begin{equation} \label{factor_46}
 \small
\setlength{\arraycolsep}{1pt}
\begin{aligned} \mathbf {F_{4 \times 6}}=\left [{ 
\begin{matrix} 
0 &\quad 1 &\quad 1 &\quad 0 &\quad 1 &\quad 0 \\ 
1 &\quad 0 &\quad 1 &\quad 0 &\quad 0 &\quad 1 \\ 
0 &\quad 1 &\quad 0 &\quad 1 &\quad 0 &\quad 1 \\ 
1 &\quad 0 &\quad 0 &\quad 1 &\quad 1 &\quad 0 \\ 
\end{matrix} }\right],
\end{aligned}
\end{equation}
\begin{equation}\label{factor_510}
 \small
\setlength{\arraycolsep}{1pt}
\begin{aligned} \mathbf {F_{5 \times 10}}=\left [{ 
\begin{matrix} 
1 &\quad 1 &\quad 1 &\quad 1 &\quad 0 &\quad 0 &\quad 0 &\quad 0 &\quad 0 &\quad 0\\ 
1 &\quad 0 &\quad 0 &\quad 0 &\quad 1 &\quad 1 &\quad 1 &\quad 0 &\quad 0 &\quad 0\\ 
0 &\quad 1 &\quad 0 &\quad 0 &\quad 1 &\quad 0 &\quad 0 &\quad 1 &\quad 1 &\quad 0\\ 
0 &\quad 0 &\quad 1 &\quad 0 &\quad 0 &\quad 1 &\quad 0 &\quad 1 &\quad 0 &\quad 1\\ 
0 &\quad 0 &\quad 0 &\quad 1 &\quad 0 &\quad 0 &\quad 1 &\quad 0 &\quad 1 &\quad 1\\ 
\end{matrix} }\right].\end{aligned}
\end{equation}

In the downlink channel, $J$ users’ data are first superimposed at the base station and then transmitted over $K$ orthogonal subcarriers. The received signal of user $u$ can be expressed as
\begin{equation}  \label{downlink}
 \small
\mathbf{y}_u=\text{diag}\left( {{\mathbf{h}}_{u}} \right)\sum\limits_{j=1}^{J}{{{\mathbf{x}}_{j}}}+\mathbf{n}_u,
\end{equation} 
where ${{\mathbf{h}}_{u}}={{\left[ {{h}_{u,1}},{{h}_{u,2}},\ldots {{h}_{u,K}} \right]}^{\text{T}}} \in {{\mathbb{C}}^{K\times 1}}$ denotes Rician fading channel   vector between the BS and $u$th IoT user.    Namely, ${{h}_{u,k}} \sim \mathcal{CN}\left ({\sqrt{ \frac {\kappa}{1+\kappa}}, \sqrt{ \frac {1}{1+\kappa}}  }\right )$ with  $\kappa$ represents the ratio of average power in the  LoS path  over that in the scattered component. The probability density function of the normalized Rician random variable  is given by  \cite{xin2003space}
 \begin{equation}
 \begin{aligned} 
 \small
 f_{\left| h_{u,k}\right|}(x) = 2x (1 + \kappa) &\exp\left(-\kappa - x^{2}(1 + \kappa)\right) \\
&\times \ I_{0}\left(2x\sqrt{\kappa(1 + \kappa)}\right), \quad x \geq 0,
 \end{aligned}
 \notag
 \end{equation}
where $I_{0}(\cdot)$ is the first order modified Bessel function of the first kind.   For simplicity of analysis, we assume the $K$ subcarriers follow the same Rician distribution of $\kappa$.  
${{\mathbf{n}_u} = \left [{ {n_{u,1},{n_{u,2}},\ldots ,{n_{u,K}}} }\right ] }^{\text{T}}$ is the complex additive white Gaussian noise   vector each element of which has zero mean and variance $N_{0}$, i.e., ${{n}_{u,k}} \sim \mathcal{CN}\left ({0, {N_0} }\right )$.  For simplicity, hereafter the subscript $u$ is omitted throughout  this paper.

\subsection{SCMA Detection}



 Similar to many previous works, we assume that the channel state information  is perfectly known by receiver \cite{mheich2018design,cai2016multi,deka2020design,yu2015optimized,Zhang,li2020design,chen2020design}. 
Thanks to  the sparsity property of the SCMA codewords, the MPA detector is  applied to reduce the decoding complexity. Define $I_{{{{r}_{k}}\to {{u}_{j}}}^{{}}}^{(t)} ({\mathbf{x}_{j}})$ as the belief message associated with codewords $\mathbf{x}_{j}$ which is transmitted from resource node ${r}_{k}$  to variable node ${u}_{j}$  at the $t$th iteration. Similar, denoted by $I_{{{u}_{j}}\to {{r}_{k}}}^{{(t)}} ({\mathbf{x}_{j}})$ the  belief message from   variable node ${u}_{j}$ to  resource node ${r}_{k}$.  We assume equal  probability for the input message $\mathbf{x}_{j}$, i.e., $I_{{{r}_{k}}\to {{u}_{j}}}^{(0)}({\mathbf{x}_{j}})=\frac{1}{M},   \forall j=1,  \ldots, J, \forall k\in \boldsymbol{\boldsymbol{\mathcal {I}}}_u(j)$. The iterative messages exchanged between  resource nodes and  variable nodes are computed as
 \begin{equation}
  \small
  \label{FN_up}
 \begin{aligned}
 I_{r_k \rightarrow u_j}^{(t)}(\mathbf {x}_j) & \\ = \sum _{\substack{\mathbf {x}_j=\mathbf {x}\\ i\in \boldsymbol{\boldsymbol{\mathcal {I}}}_r(k)\backslash \lbrace j\rbrace \\ \mathbf {x}_i \in \mathcal {X}_{i}}} &\frac{1}{\pi N_0} \text{exp} \left\{-\frac{ \vert y_k -  h_{k}  \sum  _{i\in \boldsymbol{\boldsymbol{\mathcal {I}}}_r(k)}x_{k,i} \vert ^2}{N_0} \right\}\\ &\times \prod _{i\in \boldsymbol{\boldsymbol{\mathcal {I}}}_r(k)\backslash \lbrace j\rbrace } I_{u_i \rightarrow r_k}^{(t -1)}(\mathbf {x}_i), 
 \end{aligned}
\end{equation}
and
\begin{equation}
 \small
  \label{VN_up}
I_{u_j \rightarrow r_k}^{(t)}(\mathbf {x}_j) =\alpha _j\times \prod _{\ell \in \boldsymbol{\mathcal {I}}_u(j)\backslash \lbrace k\rbrace } I_{r_{\ell } \rightarrow u_j}^{(t-1)}(\mathbf {x}_j), \end{equation}
where $\alpha _j$ is  a normalization factor. 

 

At each iteration, the main complexity of MPA is   dominated  by the message updating at the resource node, which can be approximated as $\mathcal{O} \left(K M^{d_f}\right)$, where $d_f$ is the number of users which collide  over a resource node.    To reduce the decoding complexity of MPA at each iteration, one possible approach is to reduce the effective $M$ for the codebook, which is the key idea  of the proposed LPCBs.

\section{proposed design criteria of sparse codebook in downlink Rician channels}
In this section, we first analyze  the PEP performance of SCMA in the downlink Rician channel. Then, we present the  corresponding  codebook design criteria. We also show that the proposed design criteria can also generalize the existing design criteria of MED and MPD   in  AWGN and Rayleigh fading channels, respectively.  

\subsection {PEP analysis in downlink Rician fading channels}

In downlink SCMA systems, users' data are first superimposed over $K$ orthogonal resources, which constitute superimposed  constellation ${\Phi }_{M^J}$.    Let $ \mathbf{w} = \sum _{j=1}^{J}{{{\mathbf{x}}_{j}}}$ be a superimposed codeword of ${{\Phi }_{ {M^J}}}$.  Assume the  erroneously decoded codeword is $\hat{\mathbf{w}}$ when  $ \mathbf{w}$ is transmitted, where  $\mathbf{w}, \hat{\mathbf{w}}  \in {{\Phi }_{M^J}}, $ and $ \hat{\mathbf{w} \neq \mathbf{w}}$. Furthermore, Let us define the element-wise distance $\tau _{{{\mathbf w}} \rightarrow {\tilde{\mathbf{w}}}}(k) =   \left|  \sum\nolimits_{j \in \boldsymbol{\mathcal {I}}_r(k)}(x_{j,k}- \hat{x}_{j,k}) \right|^2 $ and Euclidean distance $  \delta_{{{\mathbf w}} \rightarrow {\tilde{\mathbf{w}}}} = \sum _{k=1}^{K} \tau _{{{\mathbf w}} \rightarrow {\tilde{\mathbf{w}}}}(k)$. 
Then, the  PEP conditioned on  the  channel fading vector  for a maximum-likelihood receiver  is given as \cite{boutros1996good}
\begin{equation}
\label{PEP1}
 \small
 \begin{aligned}
\text{Pr} \{\mathbf{w} \to \mathbf{\hat{w}} | \mathbf{h} \} = &   Q\left(\sqrt {\frac{\left\Vert \text{diag} (\mathbf{h}) \sum\nolimits_{j = 1}^J (\mathbf{x}_j - \mathbf{\hat{x}}_j) \right\Vert^2}{2N_0}} \right) \\
= & Q\left(\sqrt {\frac{ \sum\nolimits_{k = 1}^K  h_k^2 \tau _{{{\mathbf w}} \rightarrow {\tilde{\mathbf{w}}}}(k)}{2N_0}} \right) ,
 \end{aligned}
\end{equation}
where  $Q \left( \cdot \right)$ is the Gaussian $ Q $-function i.e., $Q(x)=(2\pi)^{-1/2}\int _{x}^{+\infty }e^{-t^{2}/2}dt$. A  tight upper bound of $Q$-function can be expressed by  $ Q \left( x \right)  \leq \sum \nolimits _{i=1}^{L} a_i e^{-b_i x^2}$, where $L, a_i, b_i$ are constant.  Then,   by applying the above upper bound and taking expectation with respect to $\mathbf h$ on both sides of (\ref{PEP1}), we arrive
  \begin{equation}
  \label{PEP3}
   \small
\begin{aligned} 
&\text{Pr}  \{\mathbf{w} \to \mathbf{\hat{w}}  \}   \leq  \mathbb E_{\mathbf h} \left [ \sum \limits _{i=1}^{L} a_i{\prod \limits _{k = 1}^K {  \exp \left({ -\frac{{ { b_i h_k^2 { \tau _{{{\mathbf w}} \rightarrow {\tilde{\mathbf{w}}}}(k) } }}}{{ {2N_0}}}} \right) } }\right ]\\
 & =   \sum \limits _{i=1}^{L} a_i{\prod \limits _{k = 1}^K { \mathbb E_{\mathbf h} \left [ \exp \left({ -\frac{{ { b_i h_k^2 { \tau _{{{\mathbf w}} \rightarrow {\tilde{\mathbf{w}}}}(k) } }}}{{ {2N_0}}}} \right)\right ] } }\\
 &  \overset{(\mathrm i)}{=}    \sum \limits _{i=1}^{L} a_i  \Bigg \{ \prod \limits _{k = 1}^K 
\frac{1+\kappa}{1\!+ \! \kappa \!+\!\frac{{ b_i \tau _{{{\mathbf w}} \rightarrow {\tilde{\mathbf{w}}}}(k)  }  }{2N_0} }    {  {\exp \left({ \frac{- \kappa  \frac{b_i \tau _{{{\mathbf w}} \rightarrow {\tilde{\mathbf{w}}}}(k)}{2N_0}   }{1\!+ \! \kappa \!+\!\frac{{  b_i \tau _{{{\mathbf w}} \rightarrow {\tilde{\mathbf{w}}}}(k)} }{2N_0} }} \right) } }\Bigg \}, 
\end{aligned}
  \end{equation}
where step (i) is derived based on the fact that  the distribution of $h_k^2$ relates to the non-central chi-square distribution with its  moment  generating function   given by \cite{TellamburaEvaluation} 
\begin{equation}
 \small
 \label{momo}
M_{|h_k |^2} ( s) = \frac{1+\kappa}{1 +\kappa+ s}\exp \left(- \frac{\kappa s}{1 +\kappa + s} \right).
\end{equation}

By choosing $L =1$, $a_1 = b_1 =\frac{1}{2}$ we can obtain the Chernoff bound \footnote{Alternatively, one can also use $L =2, a_1 =\frac{1}{12},  b_1 =\frac{1}{2},a_2 =\frac{1}{4},  b_2 =\frac{2}{3}$ for the approximation.  In general, the  Chernoff bound may be a little loose compared to the approximation of $L \geq 2$. However, this does not affect the codebook optimization \cite{li2020design,yu2015optimized,chen2020design}. }. Accordingly, the PEP can be written in the form 
\begin{equation}
 \small
 \label{PEP_Raician2}
\begin{aligned} 
\text{Pr} \{\mathbf{w} \to \mathbf{\hat{w}}\}   \leq  \frac{1}{2} \exp \left( -{\frac{d_{ \mathbf{w} \to \mathbf{\hat{w}} }^2}{4N_0}}\right),
\end{aligned}
  \end{equation}
with 
\begin{equation}
 \small
 \label{d}
\begin{aligned} 
d_{ \mathbf{w} \to \mathbf{\hat{w}} }^2 = \sum \limits _{k=1}^K   &  \Bigg\{  {\underbrace{ { \frac{ \kappa {\tau _{{{\mathbf w}} \rightarrow {\tilde{\mathbf{w}}}}(k)}   }{1+ \kappa +\frac{{  \tau _{{{\mathbf w}} \rightarrow {\tilde{\mathbf{w}}}}(k)} }{4N_0} }    }  }_{d_{1, \mathbf{w} \to \mathbf{\hat{w}} }^2(k)} } \\ 
  & \; \; \; \; {+  \underbrace{ 4N_0   \ln \left(1+  \frac{{  \tau _{{{\mathbf w}} \rightarrow {\tilde{\mathbf{w}}}}(k)  }  }{4N_0 \left( 1+ \kappa\right)}   \right)}_{d_{2, \mathbf{w} \to \mathbf{\hat{w}} }^2(k)}  } \Bigg\}. 
\end{aligned}
  \end{equation}

     \subsection{Design criteria of multi-dimensional codebooks}    
  \label{designC}
    Define $n_e\left(\mathbf {w},\hat {\mathbf {w}}\right)$  as the erroneous bits when $\hat {\mathbf {w}}$ is decoded at receiver, the union bound of  average bit error rate (ABER) for SCMA systems is given below:
  \begin{equation}
  \label{ABER}
  \small
  \begin{aligned} P_{\text {b}} \leq \frac {1}{M^{J} \cdot J\log _{2}(M)}   \sum _{\mathbf {w}}\sum _{\hat {\mathbf {w}}\neq \mathbf {w}}{n_{\text {E}}(\mathbf {w},\hat {\mathbf {w}})}   \text{Pr} \{\mathbf{w} \to \mathbf{\hat{w}}\}. \\
  \end{aligned}
    \end{equation}
 
  

At large SNRs, as the ABER is mainly dominated  by the largest value of PEP in (\ref{PEP_Raician2}),  we assert that improving  the minimum distance of $d_{ \mathbf{w} \to \mathbf{\hat{w}} }^2$ among all pairs    will lead to lower  ABER. 
Hence, we formulate our codebook design of  the SCMA system  with structure ${\cal S}({\cal V},{\cal G}; J, M, N, K), \, \ {\cal V}: =[{\bf V}_{j}]_{j=1}^{J}$ and ${\cal G}:=[g_{j}]_{j=1}^{J}$ in downlink Rician fading channels as 
 \begin{equation}
 \small
 \label{Opt1}
 \begin{aligned}
  \mathcal {P}_1: \quad   {\cal V}^{\ast},{\cal G}^{\ast}= & \quad \quad \arg\max\limits_{{\cal V},{\cal G}}  \; {{\Delta}_{\min} \left( { \boldsymbol {\mathcal X}} \right)},\\
     \text{Subject to } & \quad \quad  \begin{array}{l}
      \sum\nolimits_{j=1}^{{J}}{ \boldsymbol {\mathcal X}_j} = MJ,
        \end{array}
      \end{aligned}
 \end{equation}
 where $\boldsymbol {\mathcal X}_j, j=1,2,\ldots,J$ are generated based on (\ref{scmaMapping}), $\boldsymbol {\mathcal X} = \{\boldsymbol {\mathcal X}_1,\boldsymbol {\mathcal X}_2,\ldots,\boldsymbol {\mathcal X}_J\}$ and  ${{\Delta}_{\min}\left( { \boldsymbol {\mathcal X}} \right)}$ denotes the minimum distance of the codebook $ { \boldsymbol {\mathcal X}}$ in Rician channels, which  is obtained by calculating  ${{{M}^{J}}\left( {{M}^{J}}-1 \right)}/{2}\;$ mutual distances between ${{M}^{J}}$ superimposed codewords, i.e.,
 \begin{equation} 
 \small
 \label{delta1}
 \begin{aligned}
{{\Delta}_{\min}\left( { \boldsymbol {\mathcal X}} \right)}  \triangleq    \min \big\{ { d_{ \mathbf{w} \to \mathbf{\hat w} }^2  \vert \;    \forall  {{\mathbf{w}}},{{\mathbf{ \hat w}} }\in {\Phi }_{M^J}, 
     {\mathbf{ w}\ne \mathbf{\hat w} }  }   \big\}.
 \end{aligned}
  \end{equation}

 Define the multiple error events (MEEs) as  the detection errors occurred with multiple users, and the  single error event (SEE) as  the detection error  occurred with a single user. Then, we   introduce the following lemma.

 {\textit{Lemma 1:}} At sufficiently  high SNR values, for $(1+\kappa) \gg \frac{ \tau _{{{\mathbf w}} \rightarrow {\tilde{\mathbf{w}}}}(k)  }{4N_0}$, the SEE  dominate the ABER, whereas  for $(1+\kappa) \ll \frac{ \tau _{{{\mathbf w}} \rightarrow {\tilde{\mathbf{w}}}}(k)  }{4N_0}$, the MEEs  dominate the ABER. Moreover,
the  proposed design criteria can also  generalize the    previous design  metric of MED and MPD in the AWGN and Rayleigh fading channels, respectively. 

\textit{ Proof:} Refer to Appendix \ref{AppeA}.


\section{Proposed Low-Projection  Codebook Design }
This section introduces our proposed LPCBs that enable  low complexity detection by maximizing the proposed  minimum distance metric, i.e., ${{\Delta}_{\min}}$.   The main design steps are: generate a one-dimensional LP-GAM vector, permute LP-GAM to obtain an $N$-dimensional MC, optimize the   constellation operators by addressing (\ref{Opt1}), and optimize the bit-to-symbol labeling.   To enhance the shaping gain, we propose to jointly optimize the MC and  constellation operators.





\subsection{Design of  one-dimensional LP-GAM}
Denote  $\mathcal A _{{M,T}}$ as a length-$M$  one-dimensional LP-GAM vector with  $T$  distinct elements, i.e.,   $M-T$ constellation points are overlapped. To obtain  $\mathcal A _{{M,T}}$, we first construct a constellation $\mathcal{A}_T$ with $T$ points, then repeat $M-T$ points in $\mathcal{A}_T$ by following certain rules.  
The selection  of the  $\mathcal A _{T}$ is of vital importance to    downlink SCMA codebook design. 
In this paper, we employ  GAM to design the $\mathcal A _{T}$ due to the attractive features of enhanced mutual information, distance and  PAPR performance \cite{larsson2017golden}.    The $n$th constellation point of GAM with $N_{p}$ points can be   generated according to $x_n = r_ne^{i2\pi \omega n}$, where $r_n = c_{\text {norm}}\sqrt{n}$, $c_{\text {norm}} = \scriptstyle\sqrt{\frac{2P}{N_p+1}}$,  $P$ is the  power constraint and $\omega = \frac{1-\sqrt{5}}{2}$ is the golden angle in radians.   
   The design  procedure of $\mathcal A _{{M,T}}$ is specified as follows:


\subsubsection{Generate $\mathcal A _{T}$}
Generate the   GAM points  according to   $x_n = c_{\text {norm}}\sqrt{n + \rho } e^{i2\pi ( \varphi+\omega)  n}$, where $\varphi$ is an arbitrary angle and $\rho$ is the amplitude factor.  Note that in the proposed LP-GAM, we have   two additional degrees of freedom for enhanced codebook design,   i.e., $ \rho $ and $\varphi$. Such a constellation is referred to as  $(\rho,\varphi)$-GAM. The construction steps of  $\mathcal A _{T}$ are   as follows:  1) For even value of $T$,  generate $N_p = \frac{T}{2}$  $(\rho,\varphi)$-GAM points,   denoted by $\mathcal A _{T}^{(m)}, m = 1, 2, \ldots,\frac{T}{2}$. The remaining $ \frac{{T}}{2}$ points are obtained by symmetry: $\mathcal A _{T }^{(m+{\frac{T }{2}})}  = -\mathcal A _{T}^{(m)}$ for $m = 1, 2, \ldots,\frac{T}{2}$;  2) For odd value of $T$,  generate $N_p = \frac{T-1}{2}$  $(\rho,\varphi)$-GAM points. Similarly, the other  $ \frac{T-1}{2}$ points are obtained by symmetry. Finally, the zero point in the complex domain is added to $\mathcal A _{T} $.

\begin{figure}[htbp]
	\centering
	\begin{subfigure}{0.46 \textwidth}
  \includegraphics[width=1 \linewidth]{{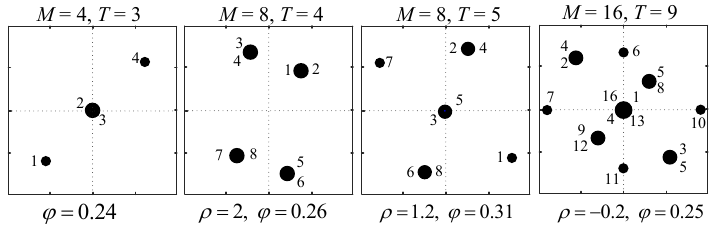}}
		\caption{One dimensional LP-GAM  }
				\vspace{-0.1em}
	\end{subfigure}
	\begin{subfigure}{0.46\textwidth}
  \includegraphics[width=1 \linewidth]{{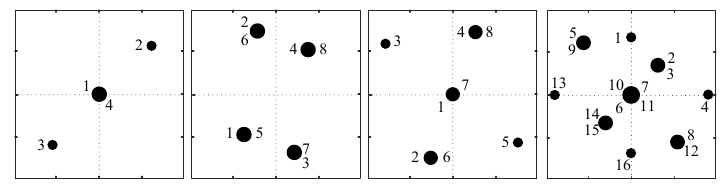}}
		\caption{Permuted LP-GAM on the second dimension  }
		\vspace{-0.1em}
	\end{subfigure}
	\caption{Examples of  $\mathcal A _{{M,T}}$ and   $ {\pi }_{2} \left( \mathcal A _{{M,T}} \right)$.}
	\label{LP-GAM}
	\vspace{-1.5em}	
\end{figure}

\subsubsection{Obtain  $\mathcal A _{{M,T}}$ based on $\mathcal A _{T}$}

After generating  the $\mathcal A _{T}$ based on GAM,  the next problem is  how to allocate  $\mathcal A _{T}$ with  overlapped signal points to a length-$M$ vector.     To proceed, let us define $\mathcal P_{M,T}= \left\{ p_1, p_2, \dots, p_L \right\}$ and $\mathcal T_{M,T}= \left\{ t_1, t_2, \dots, t_L \right\}$  as the overlapped signal set and corresponding   projected  numbers, respectively, where $p_l \in \mathcal A _{T}, 1\leq l \leq L$ and $ \sum\nolimits_{l=1}^{L}  t_l=M-T$. Namely, the constellation point $p_l$ is overlap $t_l$ times in $\mathcal A _{{M,T}}$.
The following rules are adopted to determine   $\mathcal P$ and  $\mathcal T$:

\begin{itemize}
 \item The   overlapping  number for a particular constellation point, i.e., $t_l$,   should be as small  as possible,  such that the constructed MC has a good distance property. 
 
 \item  The constellation points with lower energy in $\mathcal A _{T}$ are preferentially     added to  $\mathcal P_{M,T}$, such that the constellation $\mathcal A _{{M}}$  has a small average energy.
 
  \item     $ \mathcal P $ and   $\mathcal T $ should exhibit certain symmetry,  such that the obtained $\mathcal A _{{M}}$ is also symmetrical  and with a zero mean.
\end{itemize}

Next,  we give an example of the  constructed  LP-GAM  in Fig.  \ref{LP-GAM}(a).  One can see that  there are $M-T$ constellation points  overlapped in LP-GAM,  where the overlapped constellation set, i.e.,  $\mathcal P_{M,T}$ and $\mathcal T_{M,T}$  are predefined according to the above rules. Note that the LP-GAM of $M=4$, $T=3$ is  a special case, in which the  constellation is only determined by    $\varphi$.

\subsection{ Construction of  the $N$-dimension MC}

After obtaining  the one-dimensional constellation  $\mathcal A _{{M,T}}$, the remaining $N-1$ dimensions may be  obtained by the  repetition of  $\mathcal A _{{M,T}}$.  To obtain  a higher  gain, permutation is further applied to  $\mathcal A _{{M,T}}$.  Let   $\boldsymbol {\pi }_{n}:[1,2,\ldots, M ]^{\text{T}} \rightarrow [\pi_{n,1},\pi_{n,2},\ldots, \pi_{n,M} ]^{\text{T}} $ denote  the permutation mapping of the $n$th dimension, where $\pi_{n,m}$, $1 \leq \pi_{n,m} \leq M  $ denotes the permutation index,   and for  $m \neq l$,  $   \pi_{n,m} \neq   \pi_{n,l}$. Namely,  the $\pi_{n,m}$th constellation point in  $\mathcal A _{{M,T}}$ is permuted to the $m$th position of $\boldsymbol {\pi }_{n}(\mathcal A _{{M,T}}) $ at $n$th dimension.  Then the $N$-dimensional constellation $ \boldsymbol {\mathcal C}_{MC} = \left[\mathbf {\bar{c}}_1^{\text{T}}, \mathbf {\bar{c}}_2^{\text{T}}, \ldots, \mathbf {\bar{c}}_M^{\text{T}} \right]^{\text{T}} \in \mathbb C^{N \times K}$ can be obtained as 
\begin{equation}
 \small
     \boldsymbol {\mathcal C}_{MC} = \left[ {\pi }_{1} \left( \mathcal A _{{M,T}} \right),{\pi }_{2} \left( \mathcal A _{{M,T}} \right),\ldots,{\pi }_{N} \left( \mathcal A _{{M,T}} \right)\right]^{\text{T}}.
\end{equation}

To proceed,  we further define the $N$-dimensional Cartesian product  over the  one-dimensional   $\mathcal A _{{T}}$ as 
\begin{equation}
\small
\begin{aligned}
   \mathcal A _{{T}}^{N} & = \underbrace{ \mathcal A _{{T}} \times \mathcal A _{{T}} \times \cdots \times \mathcal A _{{T}}}_{N}  \\ & \triangleq   
     \big\{(a_{1}, a_2, \ldots, a_N)   \vert 
    \forall  a_i  \in \mathcal A _{{T}}, \forall    i \in \{1,2,\ldots,N \} \big\},
\end{aligned}
\end{equation}
 where  $``\times" $  denotes the  Cartesian product. By transforming every  element in $\mathcal A _{{T}}^{N}$ into a column vector, we obtain an $N$-dimensional matrix $\mathbf A \in \mathbb C^{N \times T^N}$. For simplicity, we denote the above  Cartesian mapping as  $ f_{\mathbf A}:  \mathcal A _{{T}}^{N} \rightarrow \mathbf A $.  
Then, we introduce the following lemma:

{\textit{Lemma 2:}} To construct an $N$-dimensional  $  \boldsymbol {\mathcal C}_{MC}$, the LP number should satisfy $\left\lceil   {\sqrt[N]{M}}\right\rceil  \leq T \leq M$. For the case  of $ {T = \sqrt[N]{M}} \in \mathbb Z$, the permutation $\pi_n, n=1,2,\ldots,N$ are given by the    mapping   $ f_{\mathbf A}$.
  
   \textit{Proof:}  Refer to Appendix \ref{AppeB}.

For the other values of $T$ and $M$, we first fix the constellation at the first dimension with a randomly chosen permutation.  Accordingly, for the second dimension, there will be a total of $T=M!$ pairing options. Consequently, there are $T= \left (M! \right)^{N-1}$  different choices of the permutations.  The  MC constrained PEP can be employed to aid the design of permutation metric. 
To this end,    we further define $d^{\boldsymbol {\mathcal C}_{MC}}_{\mathrm {min},t}, t=1,2,\ldots,T$ as the minimum effective  distance   from the $ M \left( M-1\right)/2$ mutual codeword distances   arising from   $ \boldsymbol {\mathcal C}_{MC} $.  Based on  the results in (\ref{PEP_Raician2}) and (\ref{d}), we have
\begin{equation}
\small
 \label{d_mc}
\begin{aligned} 
d^{\boldsymbol {\mathcal C}_{MC}}_{\mathrm {min},t}= \arg \underset {1 \leq i < l \leq M}{\min }       & \sum \limits _{n=1}^N   { \frac{ \kappa  \left|    c_{n,i}^{(t)}-c_{n,l}^{(t)}   \right |^2  }{1+ \! \kappa +\frac{\left|    c_{n,i}^{(t)}-c_{n,l}^{(t)}   \right |^2 }{4N_0} }    }  \\
  &+  4N_0   \ln \Bigg(1+ \! \frac{\left|    c_{n,i}^{(t)}-c_{n,l}^{(t)}   \right | ^2 }{4N_0 \left( 1+ \kappa\right)}   \Bigg),
\end{aligned}
  \end{equation}
 where $c_{n,i}^{(t)}$ denotes  the $i$th symbol at the $n$th entry of $\boldsymbol {\mathcal C}_{MC} $.    Hence, the permutation should be chosen to  maximise (\ref{d_mc}), i.e., 
 \begin{equation} 
   \small
\boldsymbol {\mathcal C}_{MC}^{\ast}=\arg \underset {t = 1,2,\ldots,T}{\max } d^{\boldsymbol {\mathcal C}_{MC}}_{\mathrm {min},t}.
\end{equation}

\textit{Lemma 3:}   By substituting $\kappa \rightarrow \infty$ (AWGN) and  $\kappa=0$ (Rayleigh fading) in (\ref{d_mc}), the equivalent distance metrics $d^{\boldsymbol {\mathcal C}_{MC}}_{\mathrm {min},t}$ are obtained  as  
\begin{equation} 
 \small
\label{dmin_mc}
d^{\boldsymbol {\mathcal C}_{MC}}_{  \mathrm {E, min},t}=\arg \underset {i\neq l}{\min }    \Arrowvert \mathbf {c}_{i}^{(t)}-\mathbf {c}_{l}^{(t)}\Arrowvert, 
\end{equation}
and
\begin{equation} 
 \small
d^{\boldsymbol {\mathcal C}_{MC}}_{\mathrm {P,min},t}=\arg \underset {i\neq l}{\min } \prod _{n \in \rho(\mathbf {c}_i, {\mathbf {c}}_l)} |c_{n,i}^{(t)}-c_{n,l}^{(t)}|^{2},
\end{equation}
respectively, where $\rho (\mathbf {c}_i, {\mathbf {c}}_l)$ denotes the set of  indices in which $ {c}_{n,i} \neq {{c}}_{n,l}$. Clearly, the MED and MPD are the corresponding permutation criteria  for AWGN  and Rayleigh fading channels, respectively \cite{chen2020design}.

For small constellation size, e.g.,  $M \leq 8$   and $N \leq 2$, one can find the optimal permutation by exhaustive search with reasonable computational complexity. However, for large   constellation size of   $\boldsymbol {\mathcal C}_{MC}$, it becomes intractable due to the huge number of permutation pairs.  To overcome such a difficulty, the binary switching algorithm (BSA)   \cite{zeger1990pseudo} can be modified to find the sub-optimal permutation pattern. Fig.  \ref{LP-GAM}(b) shows the  permuted constellation corresponding to the LP-GAM in  Fig. \ref{LP-GAM}(a) for a  $2$-dimensional $\boldsymbol {\mathcal C}_{MC}$. Taking    $\boldsymbol {\mathcal C}_{MC}$ with $N=2$, $M=4$, $T=3$ as an example, the second and third points are overlapped at the first  dimension; however, they are  separated at the second dimension with the permutation.  

 \subsection{Sparse codebook construction}

After   the design of the   $\boldsymbol {\mathcal C}_{MC}$,  phase rotations  are applied to generate multiple sparse codebooks.      In addition,  aiming at increasing the power diversity between users and improving the minimum distance properties   of the superimposed codewords, we introduce energy scaling to each codebook dimension   \cite{li2020design}. Specifically, the $j$th user's codebook with non-zero elements is generated by  $ \boldsymbol {\mathcal C}_{j} = \mathbf{E}_j \mathbf{R}_j  \boldsymbol {\mathcal C}_{MC}$, where $\mathbf{R}_j$ and $\mathbf{E}_j$ denote  the phase rotation matrix and power scaling matrix for the $j$th user, respectively. Note that the phase rotation matrix and power scaling matrix can be combined together to form a constellation operator matrix, denoted as $\mathbf{\Lambda}_j =\mathbf{E}_j \mathbf{R}_j$. For example, for a $2$-dimensional $\boldsymbol {\mathcal C}_{MC}$, $\mathbf{E}_j$ and $\mathbf{R}_j $ are given as \begin{equation}
 \begin{aligned}
 \small
 \label{delta}
 \mathbf{E}_j=\left[ \begin{matrix}
   E_1   &  0 \\
     0   &    E_2
\end{matrix} \right],
\mathbf{R}_j=\left[ \begin{matrix}
   e^{j \theta _1}   &  0 \\
     0   &     e^{j \theta _2} 
\end{matrix} \right],
\mathbf{\Lambda}_j=\left[ \begin{matrix}
   z_1   &  0 \\
     0   &    z_2
\end{matrix} \right],
 \end{aligned}
\end{equation}
   where $z_i = E_ie^{j \theta_i}, \forall  E_i >0,  1\leq i \leq N$. Based on $\mathbf{V}_j$  which has been introduced in Section \ref{sec2},  the $j$th user's codebook   is generated by $\boldsymbol{\mathcal X}_{j} = \mathbf {V}_{j} \mathbf{\Lambda}_j \boldsymbol {\mathcal C}_{MC}$.  We further combine the constellation operation matrix $\mathbf{\Lambda}_j$  and mapping matrix $\mathbf {V}_{j}$ together, i.e., $\mathbf{z}_{N \times J}^j =\mathbf {V}_{j} \mathbf{\Lambda}_j \mathbf{I}_K$, where $\mathbf{I}_K$ denotes a column vector  of $K 1$'s. Hence, $J$ codebooks  can be represented by the signature matrix  $\mathbf{Z}_{N \times J}= \left[ \mathbf{z}_{N \times J}^1, \dots, \mathbf{z}_{N \times J}^J \right]$.

Owing  to the sparsity of the factor graph matrix, the number of users superimposed on one resource node is $d_f $ which is less than the number of users. Therefore,  ${{d}_{f}}$ distinct rotation angles and scaling factors   of each dimension should be optimized  to distinguish the superimposed codewords. Different from the Latin rectangular   construction, we consider  the signature matrix  $ {\mathbf{Z}}_{K \times J}$ for $(4 \times 6)$ and $(5 \times 10)$ SCMA systems   \cite{li2020design} as follows: 
 \begin{equation} 
 \small
 \label{signature_46}
 {{\mathbf{Z}}_{4\times 6}}=\left[ \begin{matrix}
   0 & z_1 & z _2 & 0 & z _3 & 0  \\
   z _1 & 0 & z _2 & 0 & 0 & z _3 \\
   0 & z _3& 0 & z _2 & 0 & z _1  \\
   z _3& 0 & 0 & z _2 & z _1 & 0  \\
\end{matrix} \right],
  \end{equation}
\begin{equation}\label{signature_510}
\small
\setlength{\arraycolsep}{3.5pt}
\begin{aligned} 
 {\mathbf {Z}_{5 \times 10}}=\left [{ 
\begin{matrix} 
z _1  & z _2 & z _3& z _4 & 0 & 0 & 0 & 0 & 0 & 0\\ 
z _4 & 0 & 0 & 0 & z _1 & z _2 & z _3& 0 & 0 & 0\\ 
0 & z _3& 0 & 0 & z _4 & 0 & 0 & z _1 & z _2 & 0\\ 
0 & 0 & z _2 & 0 & 0 & z _3& 0 & z _4 & 0 & z _1\\ 
0 & 0 & 0 & z _1 & 0 & 0 & z _2 & 0 & z _3& z _4\\ 
\end{matrix} }\right].\end{aligned}
\end{equation}
 
Observing the power diversity, row-wise or column wise, with $\left | {z_{i}}\right | \ne \left | {{z}_{j}}\right |,\forall 1\le i<j\le {{d}_{f}}$,   may  help   improve the $ {{\Delta}_{\min}}$. Power imbalance among different users is also introduced into  ${{\mathbf{Z}}_{4\times 6}}$ and ${{\mathbf{Z}}_{5\times 10}}$  by $\left | {{z}_{1}}\right | +\left | {{z}_{3}}\right | \ne 2\left | {{z}_{2}}\right | $ and $\left | {{z}_{1}}\right | +\left | {{z}_{4}}\right | \ne \left | {{z}_{2}}\right | +  \left | {{z}_{3}}\right | $, respectively.   
  The phase rotation angles  $\boldsymbol{\theta} ={{\left[  
   {{\theta }_{1}},  {{\theta }_{2}}, \ldots, {{\theta }_{d_{f}}}  \right]}^{\text{T}}}$ and the energy factors  $\mathbf{E} ={{\left[  
   {{E }_{1}},  {{E }_{2}}, \ldots, {{E }_{d_{f}}}  \right]}^{\text{T}}}$ are the parameters to be optimized, where the degree of freedom for $\boldsymbol{\theta}$ and $\mathbf{E}$ is $d_{f} -1$.  The other two parameters to be jointly optimized are the $ \rho$ and $\varphi $ of the GAM constellation.  Hence, the number of   parameters  to be optimized for the  proposed codebooks   is $2 \left (  d_{f} -1  \right ) +1$ and $2 \left (  d_{f} -1  \right ) +2$ for $M=4$ and $M>4$, respectively.

Let $ \boldsymbol {\mathcal X}_{\lambda, M, T} $ be the $M$-ary codebook set  with  $T$ projected numbers  to be optimized for the SCMA system with the overloading  $\lambda \in \{ 150\%, 200\% \}$. As discussed in Subsection \ref{designC}, the design goal is to maximize the minimum distance metric    $\Delta_{\min}$   which is defined in  (\ref{delta1}).  Hence,  based on the proposed multi-dimensional codebook construction scheme, the  problem of $ \mathcal {P}_1 $ in (\ref{delta1}) is  reformulated as
 \begin{subequations}
\label{optimization2}
 \small
\begin{align}
\mathcal {P}_2: \quad   & \underset{\mathbf{E},\boldsymbol{\theta }, \rho,\varphi  }{\mathop{\max }}\quad    \; \; \Delta_{\min}\left( {   \boldsymbol {\mathcal X}_{\lambda,M,T}}\right) \tag{\ref{optimization2}}\\
\text {Subject to}  \quad \quad   
&\sum\limits_{i=1}^{{{d}_{f}}}{{{E}_{i}}}=\frac{MJ}{K},\;  {{E}_{i}} >0, \label{optimization2:a}\\
& 0 \le {{\theta }_{i}}\le \pi, \label{optimization2:b}\\
 &     -1  < \rho \le T, \label{optimization2:c}\\
&   0 \le  \varphi \le {\pi}/2, \label{optimization2:d}\\
 & \forall i=1,2,\ldots ,{{d}_{f}}.
\end{align}
\end{subequations}

Unfortunately, it is quite difficult, if not impossible, to directly solve
the optimization problem $ \mathcal {P}_2$ due to the non-convex and high computation complexity of the objective function.   For the $M=4$ and $J=6$ SCMA system, the computational complexity of $ \Delta_{\min}\left( {   \boldsymbol {\mathcal X}_{\lambda,M,T}      }\right)$ is moderate. However, as the increase of $K,J$ and $M$, it brings overwhelming computational complexity to calculate $ \Delta_{\min}\left( {   \boldsymbol {\mathcal X}_{\lambda,M,T}      }\right)$. Hence, for a highly overloaded SCMA system or high rate ($M \geq 8$) codebooks,   we   transform $ \mathcal {P}_2$ to a  feasible problem  with affordable  complexity by considering the distinct features of LPCBs. 

\textit{ 1) The case of $T={\sqrt[N]{M}} $:}   Let $\boldsymbol{\mathcal S}_{\text{sum}}^{k}$ denote   the superimposed constellation  at the $k$th resource node after removing the overlapped constellation in    $\boldsymbol {\mathcal C}_{MC}$. Namely,  all constellation points in  $ \boldsymbol{\mathcal S}_{\text{sum}}^{k}$ are unique and $\vert  \boldsymbol{\mathcal S}_{\text{sum}}^{k} \vert = T^{d_f}$. According to  (\ref{signature_46}) and (\ref{signature_510}),  $\boldsymbol{\mathcal S}_{\text{sum}}^{k}$  can be obtained by  $\boldsymbol{\mathcal S}_{\text{sum}}^{(k)} = \left\{ z_1a_1 + z_2a_2 + \ldots  + z_{d_f}a_{d_f} \vert \forall a_i\in  \mathcal{A}_{T}, i =1,2,\ldots,d_f \right\}$.
For simplicity, and since each resource node has the same superimposed constellation, the superscript $(k)$ is omitted.
Then, we introduce the following Lemma:

\textit{Lemma 4:}  For the  $ \boldsymbol {\mathcal X}_{\lambda, M, T}$ where $T= {\sqrt[N]{M}}$, solving the problem of $ \mathcal {P}_2$ by 
maximizing  $\Delta_{\min}\left( {   \boldsymbol {\mathcal X}_{\lambda,M,T}}\right) $ is equivalent to address  the following problem: 
\begin{equation}
\label{optimization3}
 \small
\begin{aligned}
\mathcal {P}_{2-1}: \quad   & \underset{\mathbf{E},\boldsymbol{\theta }, \rho,\varphi  }{\mathop{\max }}\quad    \; \; \text{MED} \big ({\boldsymbol{\mathcal S}}_{\text{sum}}  \big)  \\
\text {Subject to}  \quad \quad    &  T= {\sqrt[N]{M}}, T \in \mathbb Z, \\ & (\ref{optimization2}\text{a}),(\ref{optimization2}\text{b}),(\ref{optimization2}\text{c}),(\ref{optimization2}\text{d}),(\ref{optimization2}\text{e}),
\end{aligned}
\end{equation}
where
 \begin{equation} 
 \small
 \label{delta1}
 \begin{aligned}
\text{MED} \big ({\boldsymbol{\mathcal S}}_{\text{sum}}  \big)  \triangleq    \min \big \{   \vert  {s}_{m} -  {s}_{n} \vert ^2,  | \; & \forall   {{{s}}_{n}},{{{s}}_{m}}\in \boldsymbol {\mathcal S}_{\text{sum}},     \\
 &    {\forall m,n\in {{Z}_{{{T}^{d_f}}}},m\ne n } \big\}.
 \end{aligned}
  \end{equation}
  
\textit{Proof: } Refer to  Appendix \ref{AppeC}.

Since ${{T}^{d_f}}$ is relatively small, the complexity of calculating  $\text{MED} \big ({\boldsymbol{\mathcal S}_{\text{sum}} } \big) $ is negligible.

\textit{ 2) For the case when $ T>{\sqrt[N]{M}}$:}  We propose to maximize the lower bound of  $\Delta_{\min}\left( {   \boldsymbol {\mathcal X}_{\lambda,M,T}      }\right) $. To proceed, let us first define 
\begin{equation} 
  \small
 \label{delta2}
 d_{1, \min}^{   \mathbf {\mathcal X}_{\lambda,M,T}}  \triangleq     \min \bigg\{   \sum \limits _{k=1}^K  d_{1, \mathbf{w} \to \mathbf{\hat w} }^2  \vert \;    \forall   {{\mathbf{w}}},{{\mathbf{ \hat w}} }\in {\Phi }_{M^J}, 
     {\mathbf{ w}\ne \mathbf{\hat w} } \bigg\}   ,
  \end{equation}
and 
\begin{equation} 
  \small
 \label{delta3}
 \begin{aligned}
 d_{2, \min}^{   \mathbf {\mathcal X}_{\lambda,M,T}} & \triangleq     \min \bigg\{   \sum \limits _{k=1}^K  d_{2, \mathbf{w} \to \mathbf{\hat w} }^2  \vert \;    \forall  {{\mathbf{w}}},{{\mathbf{ \hat w}} }\in {\Phi }_{M^J}, 
     {\mathbf{ w}\ne \mathbf{\hat w} }  \bigg\}, \\
  &  \overset{(\mathrm i)}{=}  \underset {1 \leq j \leq J}{\min}  \bigg\{ \sum \limits _{k=1}^K  { 4N_0   \ln \Big(1\!  +  \! \frac{{ \left| x_{j,i,k} \!  -  \! x_{j,l,k}\right|^2   }  }{4N_0 \left(  1+    \kappa\right)}   \Big)}  \\
& \quad \quad  \quad    | \forall   x_{j,i,k}, x_{j,l,k} \in \mathbf {\mathcal X}_j ,   \forall i,l\in  {Z}_{M} ,   i \neq l \bigg\},
 \end{aligned}
  \end{equation}
where $   d_{1, \mathbf{w} \to \mathbf{\hat w} }^2$ and  $   d_{2, \mathbf{w} \to \mathbf{\hat w} }^2$ are given in (\ref{d}).   Step (i) holds true since the minimum distance metric between superimposed codewords   is achieved when the SEE occurs.  Obviously, for the  codebook  ${   \boldsymbol {\mathcal X}_{\lambda,M,T}}$,  
\begin{equation}
\small
     \Delta_{\text{LB}}\left( {   \boldsymbol {\mathcal X}_{\lambda,M,T}      }\right)  \triangleq d_{1, \min}^{   \mathbf {\mathcal X}_{\lambda,M,T}} + d_{2, \min}^{   \mathbf {\mathcal X}_{\lambda,M,T}},
\end{equation}
is a lower bound of  $\Delta_{\min}\left( {   \boldsymbol {\mathcal X}_{\lambda,M,T}      }\right) $.   Note that   $\Delta_{\text{LB}}\left( {   \boldsymbol {\mathcal X}_{\lambda,M,T}      }\right)$  and  $\Delta_{\min}\left( {   \boldsymbol {\mathcal X}_{\lambda,M,T}      }\right) $ may not achieve the minimum value at the same   solution, however,  the value of $\Delta_{\min}\left( {   \boldsymbol {\mathcal X}_{\lambda,M,T}      }\right) $ can still be improved by maximizing  $\Delta_{\text{LB}}\left( {   \boldsymbol {\mathcal X}_{\lambda,M,T}      }\right)$.
Moreover,  upon taking  into account of  lemma $1$ in Subsection \ref{designC}, one can  see that improving the $d_{1, \min}^{   \mathbf {\mathcal X}_{\lambda,M,T}}$ and $d_{2, \min}^{   \mathbf {\mathcal X}_{\lambda,M,T}}$ can also reduce the probability of MEEs and SEE, respectively. In fact, for large or small value of $\kappa$, $\Delta_{\text{LB}}\left( {   \boldsymbol {\mathcal X}_{\lambda,M,T}}\right)$ is very tight.  
Hence, instead of directly maximizing $\Delta_{\min}\left( {   \boldsymbol {\mathcal X}_{\lambda,M,T}}\right) $ in  $ \mathcal {P}_2$, we propose to address its lower bound, i.e.,
\begin{equation}
\label{optimization4}
 \small
\begin{aligned}
\mathcal {P}_{2-2}: \quad   & \underset{\mathbf{E},\boldsymbol{\theta }, \rho,\varphi  }{\mathop{\max }}\quad    \; \; \Delta_{\text{LB}}\left( {   \boldsymbol {\mathcal X}_{\lambda,M,T}      }\right)  \\
\text {Subject to}  \quad \quad    &  T > {\sqrt[N]{M}},  T \in \mathbb Z, \\ & (\ref{optimization2}\text{a}),(\ref{optimization2}\text{b}),(\ref{optimization2}\text{c}),(\ref{optimization2}\text{d}),(\ref{optimization2}\text{e}).
\end{aligned}
\end{equation}

Note that the computational complexity of $d_{2, \min}^{   \mathbf {\mathcal X}_{\lambda,M,T}}$ is negligible  since it only involves the calculation of $\boldsymbol {\mathcal X}_j$. Hence, the computational complexity of  $\Delta_{\text{LB}}\left( {   \boldsymbol {\mathcal X}_{\lambda,M,T}      }\right)$ is largely reduced compared to that of   $\Delta_{\min}\left( {   \boldsymbol {\mathcal X}_{\lambda,M,T}}\right)$. 

For the sparse codebook optimization problem, it is designed with the highly non-linear  objective function  and linear constraints. Such problem can be efficiently solved with an optimization solver, e.g. MATLAB Global Optimization toolbox. In particular, we  employ  the \textit{fmincon} solver to address the problem (\ref{optimization3}) and (\ref{optimization4}).   The maximum iteration number  for $\lambda = 150\%$ and $\lambda = 200\%$ SCMA systems are $30$ and $25$, respectively. Although the complexity of $\Delta_{\text{LB}}\left( {   \boldsymbol {\mathcal X}_{\lambda,M,T}}\right)$ has been significantly reduced,  the calculation of $ d_{1, \min}^{   \mathbf {\mathcal X}_{\lambda,M,T}}$ is still a challenging problem for $M \geq 16$ or $d_f=4$. To solve this problem,  a sub-optimal Monte Carlo search method is adopted.  The  distance metric  $ d_{1, \min}^{   \mathbf {\mathcal X}_{\lambda,M,T}}$ is first calculated between $Q$ elements, which is randomly selected form the superimposed codewords  ${{\Phi }}$. Then, the above process is repeated with $t_{\text{max}}$ times to estimate the  $ d_{1, \min}^{   \mathbf {\mathcal X}_{\lambda,M,T}}$.

\subsection{Labeling design for each  user}

 \begin{algorithm}[t] 
\caption{Modified BSA for SCMA codebook.}
\label{labelingAlg}
\begin{algorithmic}[1]
\REQUIRE {Optimized codebook $\boldsymbol{\mathcal X}_{j} $} \\
\ENSURE {A locally optimum bit-to-symbol mapping $\mathbf z$ for  $\boldsymbol{\mathcal X}_{j} $   } \\
\STATE  \textbf{Initialize}: Randomly choose an index vector $\mathbf z$    for codebook  $\boldsymbol{\mathcal X}_{j} $ and sort $\boldsymbol{\mathcal X}_{j} $  out based on the random index vector, and set $ \mathbf {z^\ast} = \mathbf {z} $ \\
\FOR {$I =1:I_{\text{max}}$ } 
\STATE { Obtain  $\Xi (\mathbf {x}_{j,i}), i=1,2,\ldots M$ and   $\Pi_{\text{ini}}$ by  updating the cost of  $\boldsymbol{\mathcal X}_{j}\left(\mathbf {z^\ast} \right) $  using  (\ref{ricianLabeling})  }
\STATE {  Sort   $\Xi (\mathbf {x}_{j,i})$   in deceasing order and obtain the new index vector $ \mathbf{z} $}
\FOR {$i=1:M$}  
\FOR {$l=1:M$}  
\STATE {\textbf{if} {$\mathbf z(i) =  l$} \textbf{then} next $l$ }
\STATE { Switch the $\mathbf z(i) $-th codeword with $l$-th codeword and update the index vector $\mathbf { \bar{z}}$  }
\STATE { Calculate the total cost $\Pi_{\text{sw}}$ using  (\ref{ricianLabeling})  }
\IF{$\Pi_{\text{sw}} \le \Pi_{\text{ini}} $}
\STATE { Retain the switch, and update  $ \mathbf {z^\ast}$ and $  \Pi_{\text{ini}} $ with $\mathbf {\bar{z}}$ and $ \Pi_{\text{sw}}  $, respectively  }
\ELSE
\STATE { Switch back the two symbols  }
\ENDIF
\ENDFOR
\ENDFOR
\ENDFOR
\end{algorithmic}
\end{algorithm}

For each   sparse codebook  $\boldsymbol{\mathcal X}_{j} $,  the corresponding bit-to-symbol mapping, i.e., bit labeling,   needs to be optimized.   Similar to the case of permutation, the  labeling rules should be  designed to minimize the MC constrained ABER. Under Rician fading channels, the labeling  metric is given as
\begin{equation}
 \small
\label{ricianLabeling}
\Pi (\xi _{t}) =  \sum _{i=1}^{M-1} \underbrace{    \sum _{\substack{l=1,l \neq i}}^{M} N_{i,l}(\xi _{j})\exp \left(\frac{-d_{t,i,l}}{4N_0} \right)}_{\Xi (\mathbf {x}_{j,i})},
\end{equation}
where $N_{i,l}(\xi _{t})$ denotes  the number of different labelling bits between $\mathbf {x}_{j,i} $ and  $\mathbf {x}_{j,l} $ based on the considered labelling rule $\xi _{t}$,  $\Xi (\mathbf {x}_{j,i})$ denotes the cost for the $i$th codeword of $\boldsymbol{\mathcal X}_{j}$, and $d_{t,i,l}$ is given as 
\begin{equation}
 \small
 \label{d_mc_j}
\begin{aligned} 
d_{t,i,l} =\sum \limits _{k=1}^K   &  \Bigg \{{ \frac{ \kappa  \left|    x_{ i,k}^{(t)} -x_{l,k}^{(t)}  \right |^2  }{1+ \kappa +\frac{\left|    x_{i,k}^{(t)} -x_{l,k}   \right |^2 }{4N_0} }    }  \\
  &   \; \; \; \;+  4N_0   \ln \Bigg(1+  \frac{\left|    x_{i,k}^{(t)}-x_{l,k}^{(t)}    \right |^2 }{4N_0 \left( 1+ \kappa\right)}   \Bigg) \Bigg \},
\end{aligned}
  \end{equation}
where  $x_{ i,k}^{(t)}$ represents  the element of $\mathbf {x}_{i} $ at the $k$th entry for the labeling rule $\xi_t$. 

\textit{Lemma 4:}  Obviously,   let   $\kappa \rightarrow \infty$  and 
  $\kappa=0$ in (\ref{d_mc_j}), we can also obtain the equivalent labeling metric  \cite{chen2020design}
 \begin{equation}
 \small
\Pi_{\text{A}} (\xi) = \sum _{i=1}^{M-1} \sum _{l=i+1}^{M} N_{i,l}(\xi )\exp \left(\frac{-\Arrowvert \mathbf {x}_{j,i}-\mathbf {x}_{j,l} \Arrowvert}{4N_0} \right), 
\end{equation} 
and 
\begin{equation}
 \small
\Pi_{\text{R}} (\xi)  =\sum _{i=1}^{M-1} \sum _{l=i+1}^{M} \frac {N_{i,l}(\xi)}{ \prod \limits_{k' \in \rho \left( \mathbf{x}_{j,i}, \mathbf {x}_{j,l}\right)}  \left|x_{j,i,k'}-x_{j,l,k'}\right|^{2}},
\end{equation}
 for AWGN  and  Rayleigh fading channels, respectively.

 There are    $T=M!$ labeling options for  an $M$-ary  constellation. Hence, for large modulation order $M$, the complexity of  exhaustive search is prohibitively high.   The BSA can be employed again to find the labeling solution with reasonable complexity. A modified BSA  for SCMA  codebooks is  described in \textbf{Algorithm \ref{labelingAlg}}. In  the BSA based labeling design, we first randomly initialize  the labelling bits to $M$ codewords, and calculate the cost value   $\Pi (\xi _{j})$  based on  (\ref{ricianLabeling}).  Then, starting  from the first codeword, we calculate the individual cost $\Xi (\mathbf {x}_{j,i})$ for all codewords and  sort  them out  in decreasing order.  Next, the algorithm swaps codeword index  with the highest individual cost with another  to find  a lower  total cost.      \textbf{Algorithm \ref{labelingAlg}} can be performed several times with different initial conditions to prevent it from jumping into local optimum.  \textbf{Algorithm \ref{labelingAlg}} can significantly reduce  the computation complexity of searching labelling pattern.  For each $M$-ary codebook,  the computation complexity of  exhaustive search  can be approximated by $\mathcal O \left( M !\right)$. In contrast, the  complexity of  \textbf{Algorithm \ref{labelingAlg}} can be estimated as $\mathcal O \left( I_{\text{max}} M^2\right)$. For example, the reduction in complexity levels reaches $79\%$ and $99\%$ when considering $M=8$ and $M=16$, respectively.

 \begin{algorithm}[t] 
\caption{Construction of LPCBs.}
\label{algorithm1}
\begin{algorithmic}[1]
\REQUIRE{$J$, $K$, $\mathbf V_j$, $M$, $T$, $\kappa$, $N_0$,   $\mathcal P_{M,T}$ and  $\mathcal T_{M,T}$}\\
\STATE  \textbf{Initialize}: $E_i$, $\theta_i$, $i=1,2,\ldots, d_f$,  $\rho$, $\varphi$ \\
\STATE   \textbf{Step 1} : For the given $T$, generate   $\mathcal A _{T}$ based GAM point.\\
\STATE   \textbf{Step 2} :  For the given $T$ and $M$, repeat the elements in $\mathcal A _{T}$ to obtain   $\mathcal A _{{M,T}}$ according to  $\mathcal P_{M,T}$ and  $\mathcal T_{M,T}$.\\
\STATE   \textbf{Step 3} : Perform  a permutation of  $\mathcal A _{{M,T}}$ according to the criteria given in (\ref{d_mc}) to obtain the $N$-dimensional MC, i.e.,  $\boldsymbol {\mathcal C}_{MC}$.\\
\STATE   \textbf{Step 4} : Generate the $j$th user's codebook  by $\boldsymbol{\mathcal X}_{j} = \mathbf {V}_{j} \mathbf{\Lambda}_j \boldsymbol {\mathcal C}_{MC}$.\\
\STATE   \textbf{Step 5} :  Jointly optimize the MC and  $\mathbf{\Lambda}_j$ by addressing the $\mathcal {P}_{2-1}$ or $\mathcal {P}_{2-2}$.   \\
\STATE   \textbf{Step 6} :  Perform  the bit labeling  with  \textbf{Algorithm \ref{labelingAlg}} for the optimized codebook $\boldsymbol{\mathcal X}_{j}^\ast$, and use the labeling metric given in (\ref{ricianLabeling}).
\end{algorithmic}
\end{algorithm}


To sum up, 
we have proposed an efficient codebook design with LP numbers in the downlink Rician fading channel  and the overall construction procedures are given in \textbf{ Algorithm \ref{algorithm1}}. 

\section{Numerical Results}

\begin{figure*}[htbp]
\centering
\subcaptionbox{$M=4$, $\lambda \in \left\{150\%, 200\%  \right\}$ and  $\text{E}_\text {b}/\text{N}_0 = 12$ dB \label{fig:Ex_Im}}{\includegraphics[width=0.44\textwidth]{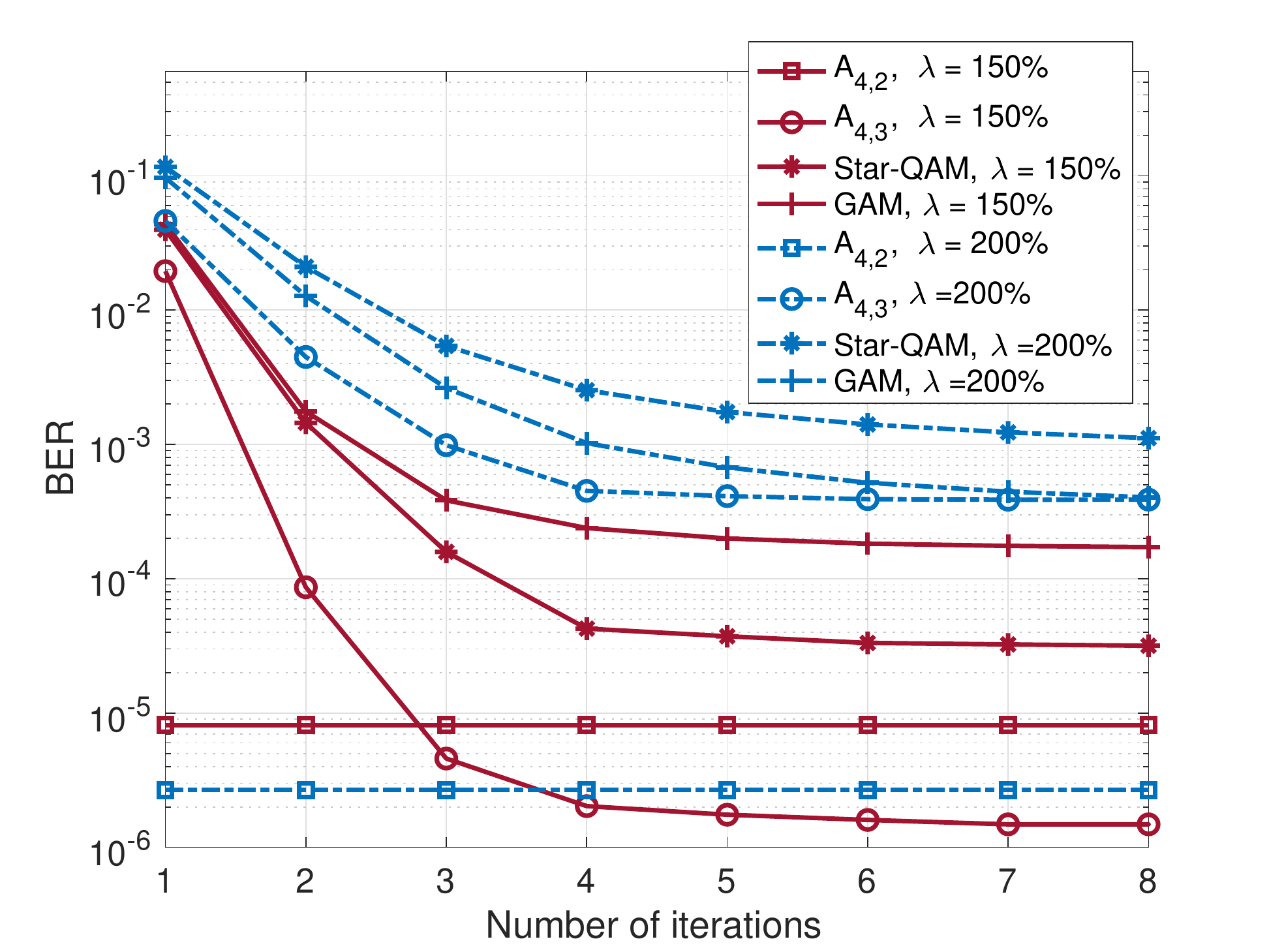}   } 
\subcaptionbox{$\lambda =150\%$, $M= 8$ ($\text{E}_\text {b}/\text{N}_0 = 14$  dB)  and $M= 16$ ($\text{E}_\text {b}/\text{N}_0 = 18$ dB) \label{fig:Ex_Im2}}{\includegraphics[width=0.44\textwidth]{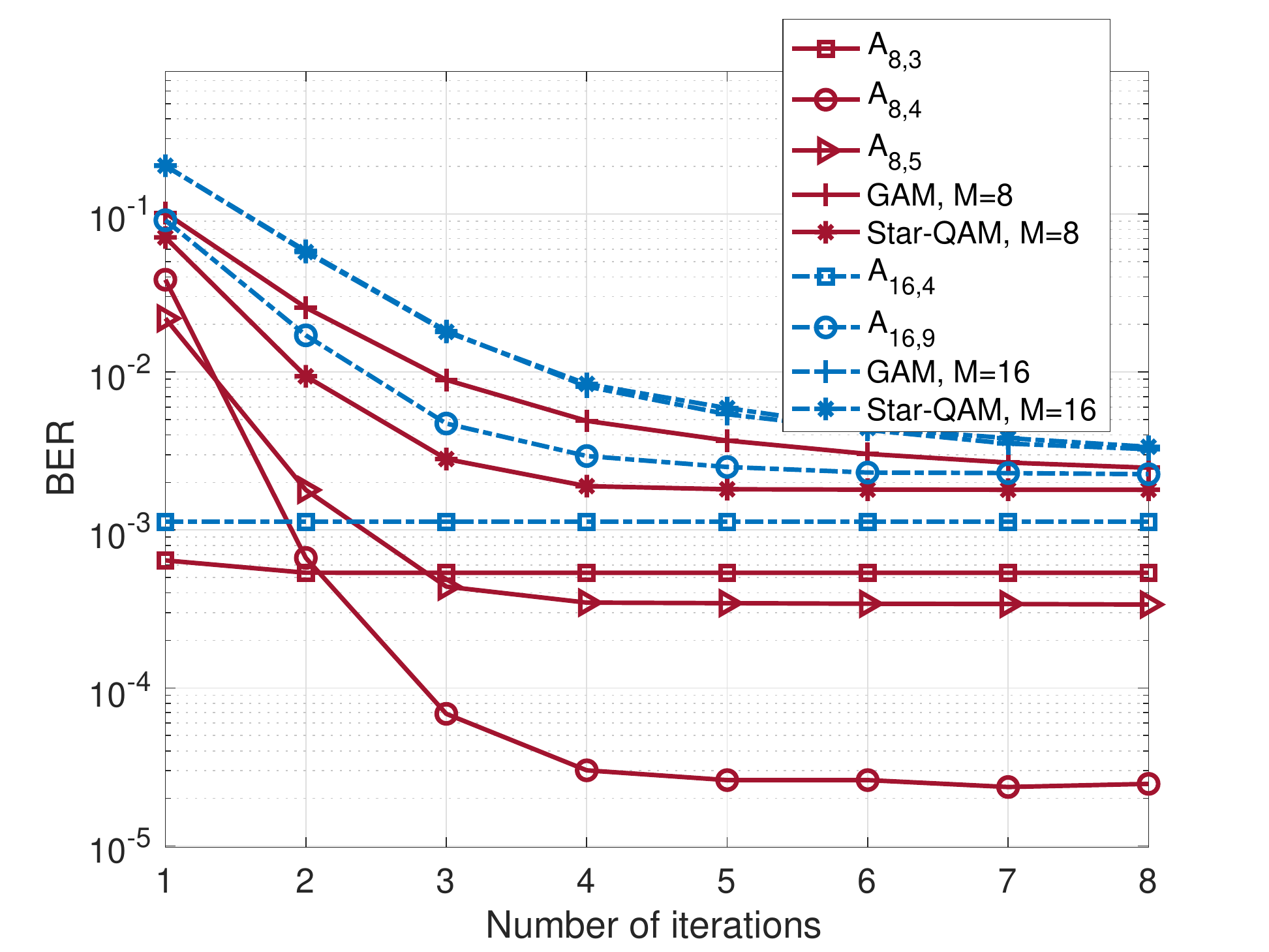} }
\caption{Number of iterations vs. BER  performance of different codebooks ($\kappa \rightarrow \infty$). } \label{MPA_iter}
\end{figure*}

In this section,  we conduct numerical evaluations of the proposed LPCBs for the $(4\times 6)$ and $(5\times 10)$ SCMA   systems characterized  in   (\ref{factor_46}) and (\ref{factor_510}), respectively.   These two systems lead to overloading factors of $\lambda =150\%$ and $\lambda = 200\%$ , respectively. 
Let $\text{A}_{M, T}$ denote  the proposed LPCB  of $M$-ary with $T$ LP numbers. Since the proposed permutation, labeling and sparse codebook optimization criteria are applicable in the   high  $\text{E}_\text{b}/\text{N}_0$ regime, we consider  $\text{E}_\text{b}/\text{N}_0 = 16$ dB, which is sufficiently large. In general, in an  IoT networks with an LoS path, a large $\kappa$ is assumed for many  scenarios \cite{5G_NR_s_iot, lutz1991land, vucetic1992channel,ErnestNOMA,WangTrajectory}. Hence, we set $\kappa = 20$ for the optimization of the proposed LPCBs. Moreover,
considering the trade-off between the computation complexity and accuracy, we set $Q=10000$ and $t_{\text{max}}=20$ for the cases of $M \geq 16$ or $\lambda = 200\%$.  The details of the designed LPCBs   are given in  our GitHub project\footnote{\url{https://github.com/ethanlq/SCMA-codebook}} and some LPCBs are also presented in Appendix C.

The main  state-of-the-art  codebooks for comparison with the proposed ones are the GAM codebook \cite{mheich2018design}  and the Star-QAM codebook \cite{yu2015optimized}\footnote{We have also optimized the MCs  that proposed  in \cite{bao2017bit} with the proposed design approach. However, it is found that the resultant BER performance is far in ferier to the GAM and Star-QAM codebooks in downlink channels. }. 
 This is because  both  \cite{mheich2018design} and \cite{yu2015optimized}   achieve  good BER performance in downlink channels. 
 We first analyze  and present the complexity reduction of our  proposed LPCBs for MPA decoder. Then, the BER performances  are  compared with  the benchmark codebooks in both uncoded and coded systems. 
  \begin{table}[]
  \small
 \centering
 \caption{Comparison of the proposed codebooks in terms of $\Delta_{\min}$,  MED and $I_t$. }
     \begin{tabular}{c|c|c|c|c}
    \hline
    \hline
       \makecell[c] {System setting \\ ($M,\lambda$)}    &  Codebooks & $\Delta_{\min}$  & MED & \makecell[c] {$I_t $  (MPA)} \\
    \hline
    \hline
    \multirow{4}{*}{$\left(4, 150\% \right)$} & $\text{A}_{4,2}$ & 1.16 & 1.1 & 1 \\
    \cline{2-5}
    &$\text{A}_{4,3}$ & 1.31 & 1.23 & \multirow{4}{*}{$ 4$}\\
    \cline{2-4}
    & Star-QAM  \cite{yu2015optimized}& 0.98 & 0.9   \\
    \cline{2-4}
     & GAM \cite{mheich2018design}   & 0.74  & 0.57  \\
   \cline{2-4}
    & Huawei \cite{huawei} & 0.67  & 0.57    \\
    \hline
    \hline
    \multirow{4}{4em}{$\left(4, 200\% \right)$} & $\text{A}_{4,2}$ & 1.05 & 0.96 & $ 1$\\
    \cline{2-5}
    &$\text{A}_{4,3}$ & 0.75 & 0.64 & $ 4$ \\
    \cline{2-5}
    & Star-QAM \cite{yu2015optimized}& 0.56 & 0.48 &\multirow{2}{*}{$ 6 $}\\
    \cline{2-4}
     & GAM \cite{mheich2018design} & 0.47 & 0.43 \\
    \hline
    \hline
    \multirow{4}{4em}{$\left(8, 150\% \right)$} & $\text{A}_{8,3}$& 0.62 & 0.54 & $2$\\
    \cline{2-5}
    &$\text{A}_{8,4}$ & 0.75 & 0.67 & $ 4$ \\
    \cline{2-5}
    &$\text{A}_{8,5}$& 0.62 & 0.53 & $ 4$ \\
    \cline{2-5}
    &$\text{A}_{8,6}$ & 0.51 & 0.46 & $ 5$ \\
    \cline{2-5}
    &$\text{A}_{8,7}$& 0.54 & 0.50 & $ 5$ \\
    \cline{2-5}
    & Star-QAM \cite{yu2015optimized}& 0.5 & 0.45   & $4$\\
    \cline{2-5}
     & GAM \cite{mheich2018design} & 0.51 & 0.47 & $6$\\
    \hline
    \hline     
    \multirow{4}{*}{$\left(16, 150\% \right)$} & $\text{A}_{16,4}$ & 0.32  & 0.26 & $ 1$\\
    \cline{2-5}
    &$\text{A}_{16,9}$&   0.28 &   0.21  & $ 5$ \\
    \cline{2-5}
    & Star-QAM \cite{yu2015optimized}& 0.2 & 0.15 &  $7 $  \\
    \cline{2-4}
     & GAM \cite{mheich2018design}  & 0.2 & 0.16 & $7 $\\
    \hline
     \end{tabular}
     \label{codebooks}
 \end{table}

Table \ref{codebooks} compares  the  $\Delta_{\min}$  of the proposed  codebooks, GAM and Star-QAM for the  SCMA  systems with overloading $\lambda =150\%$ and $\lambda = 200\%$, respectively. In addition, since the MED is widely used as a performance metric, we also present the MED of different codebooks. 
Clearly, most of the proposed LPCBs enjoy larger $\Delta_{\min}$ and MED values. 
For example,  the  MEDs of proposed codebooks $\text{A}_{4,3}$ and $\text{A}_{4,2}$ of $\lambda =150\%$ and $\lambda = 200\%$ SCMA systems, respectively,  are significantly larger than that of other codebooks. %
 It is also  worth mentioning that, to our   best knowledge, the proposed codebooks $\text{A}_{4,2}$ for  $ \lambda = 150\%$,    $\text{A}_{8,4 }$ and $\text{A}_{16,4}$ for $ \lambda = 200\%$,  enjoy the largest  MED values  among all existing codebooks.

\subsection{Complexity reduction of MPA with the  proposed LPCBs}  
In this subsection, we evaluate the  convergence behaviors of MPA  with the proposed LPCBs  and compare the receiver complexities of the proposed LPCBs with conventional codebooks  \cite{mheich2018design,yan2016top,deka2020design,yu2015optimized,klimentyev2017scma,Zhang,li2020design,chen2020design}, referred as Non-LPCBs.  

\begin{figure*}[htbp]
	\centering
	\begin{subfigure}{0.24 \textwidth}
  \includegraphics[width=1.1 \linewidth]{{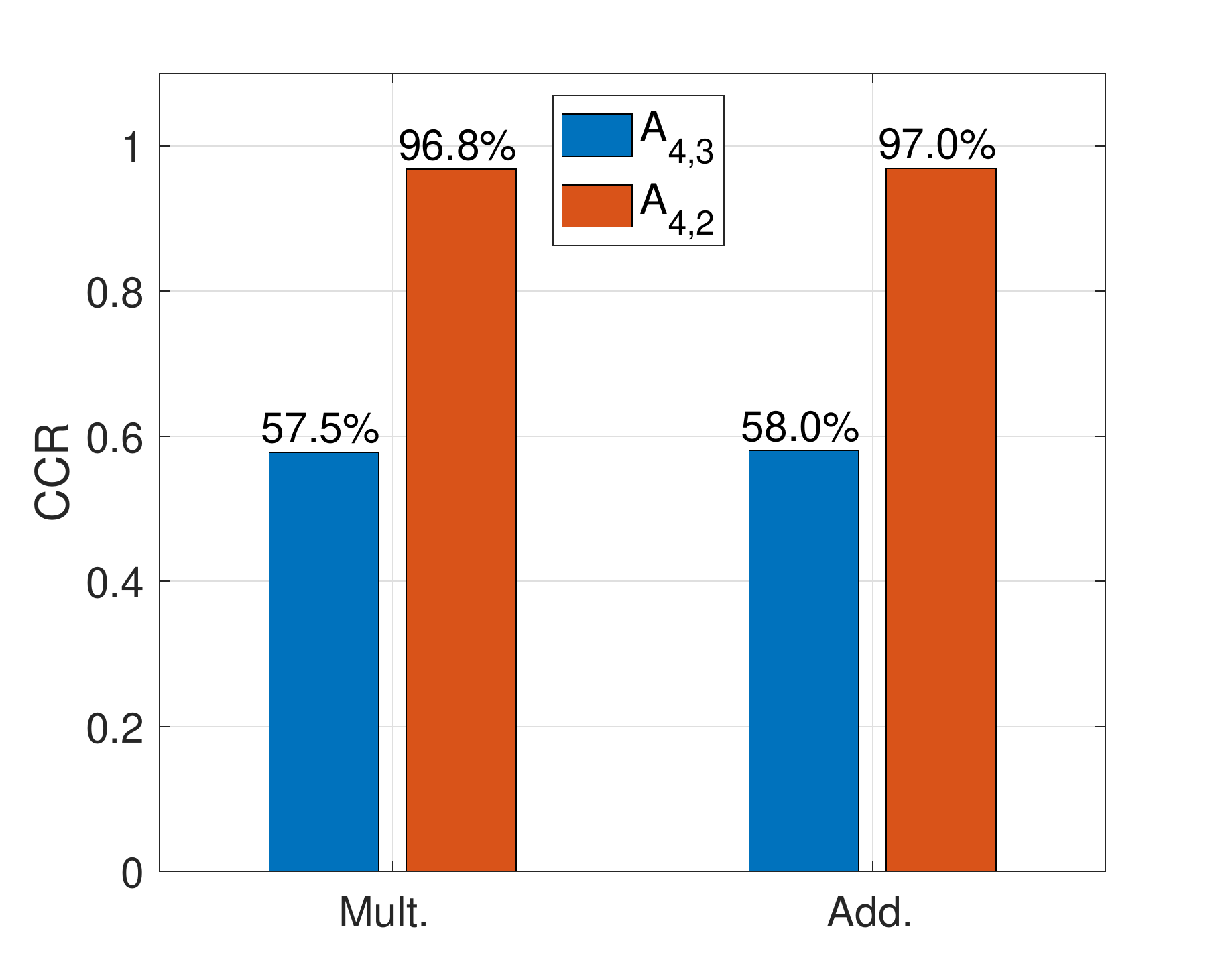}}
		\caption{$\lambda =150\%$, $M= 4$  }
				\vspace{-0.1em}
	\end{subfigure}
	\begin{subfigure}{0.24\textwidth}
  \includegraphics[width=1.1 \linewidth]{{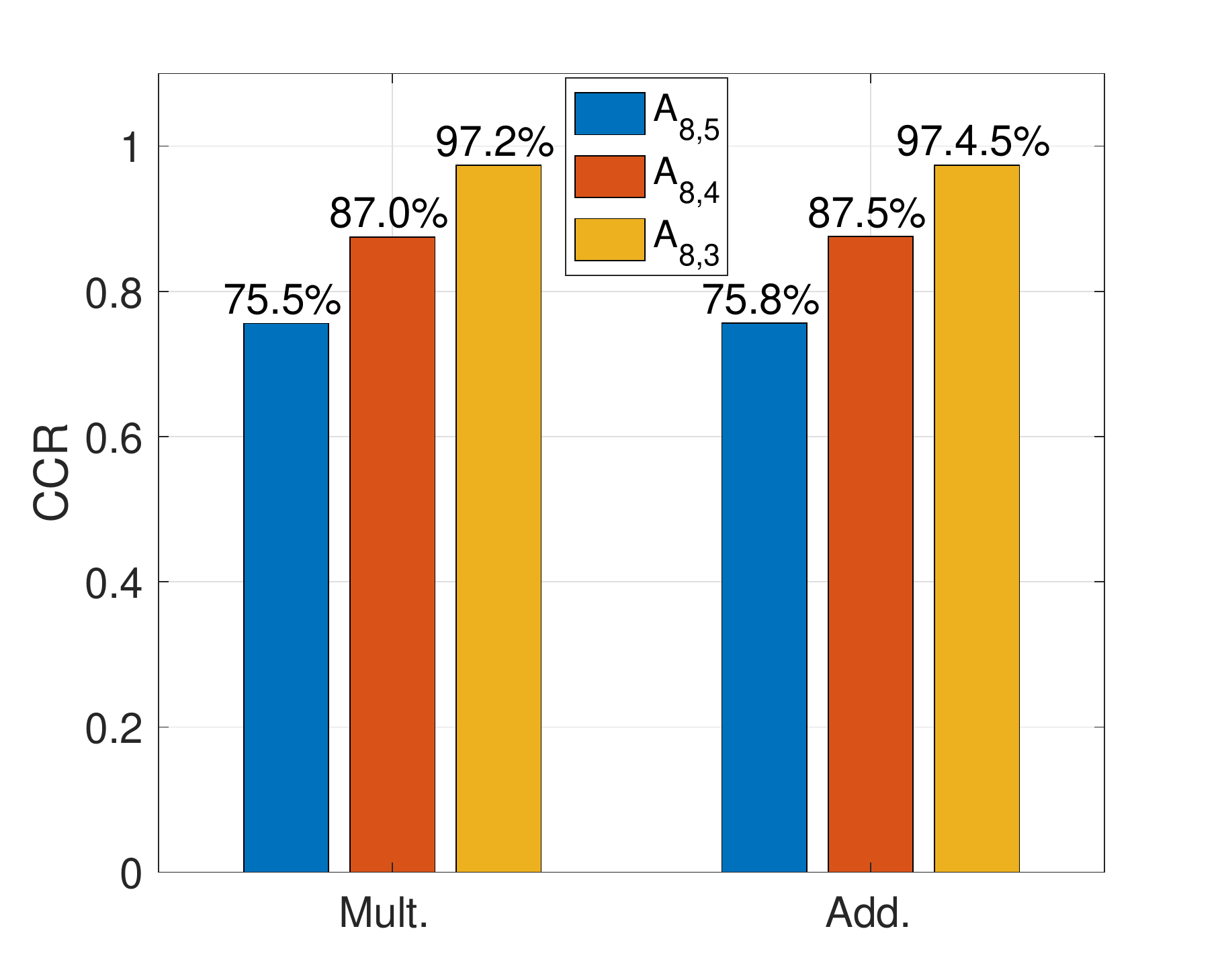}}
		\caption{$\lambda =150\%$, $M= 8$  }
		\vspace{-0.1em}
	\end{subfigure}
	\begin{subfigure}{0.24 \textwidth}
  \includegraphics[width=1.1 \linewidth]{{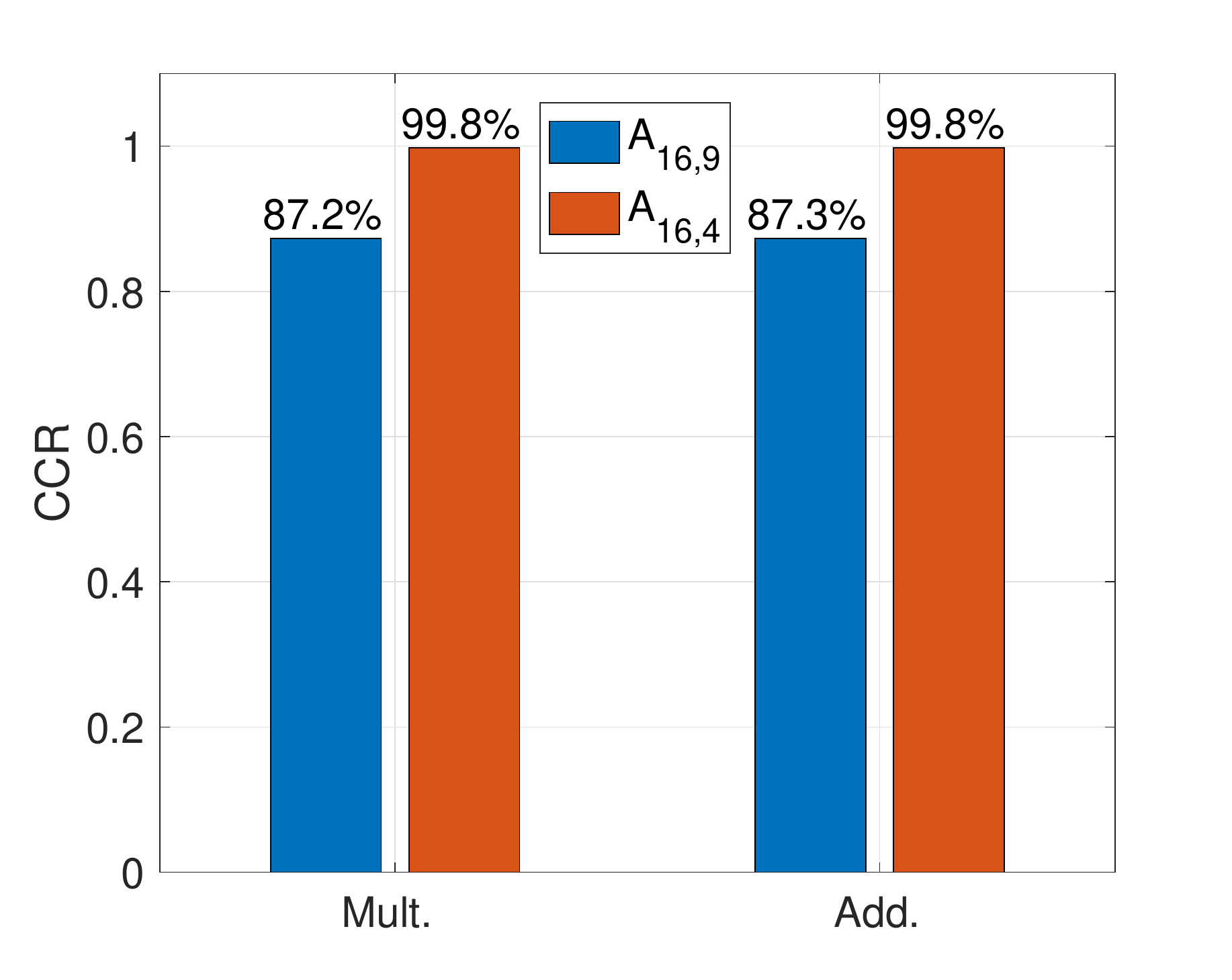}}
		\caption{$\lambda =150\%$, $M= 16$  }
				\vspace{-0.1em}
	\end{subfigure}
	\begin{subfigure}{0.24\textwidth}
  \includegraphics[width=1.1 \linewidth]{{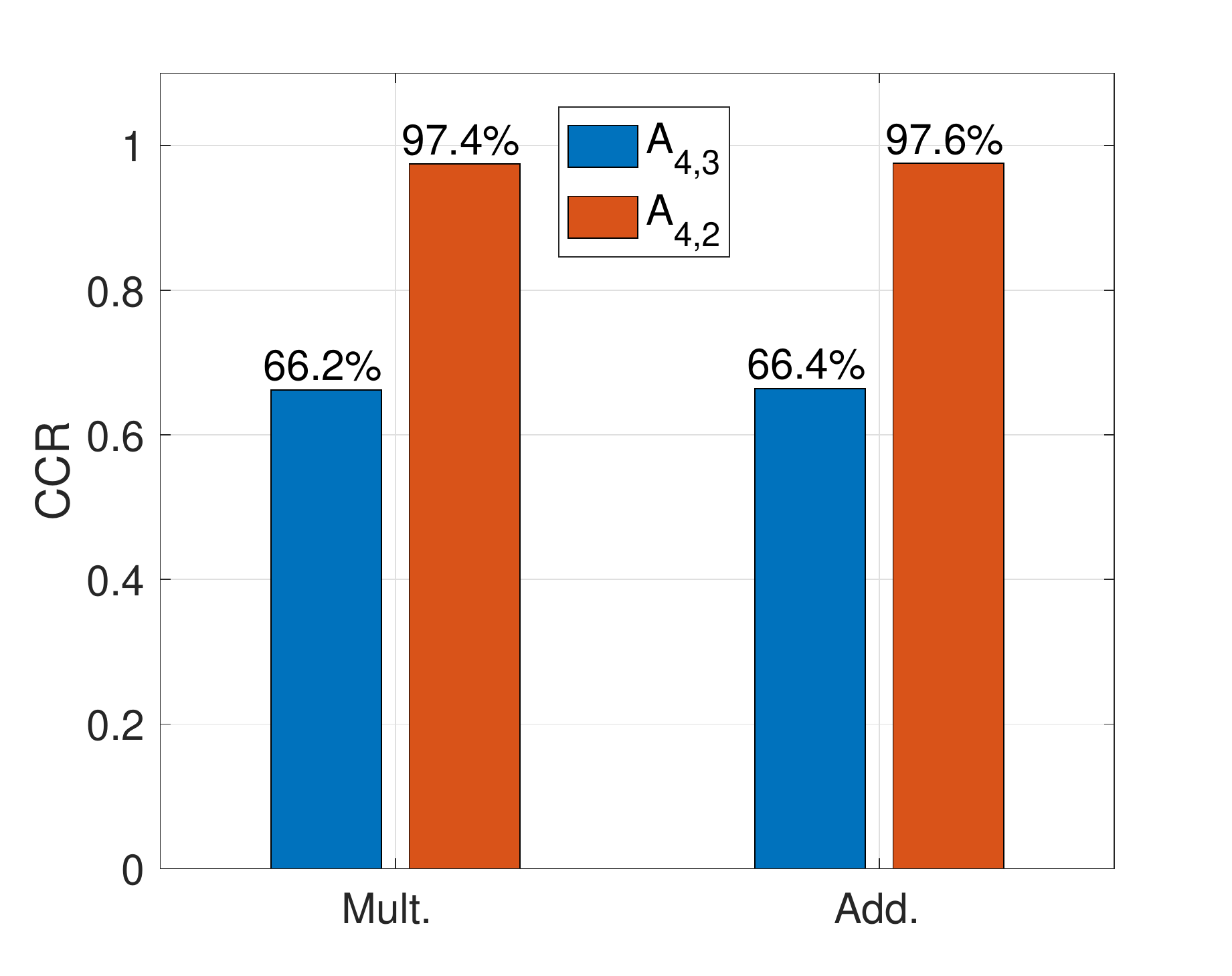}}
		\caption{$\lambda =200\%$, $M= 4$  }
		\vspace{-0.1em}
	\end{subfigure}	
	\caption{Complexity reduction of the proposed LPCBs with MPA.}
	\label{complexity}
	\vspace{-1.5em}	
\end{figure*}
 \begin{figure*} 
\centering
\subcaptionbox{$M=4$ and $\lambda \in \left\{150\%, 200\%  \right\}$ \label{fig:Ex_Im}}{\includegraphics[width=0.42\textwidth]{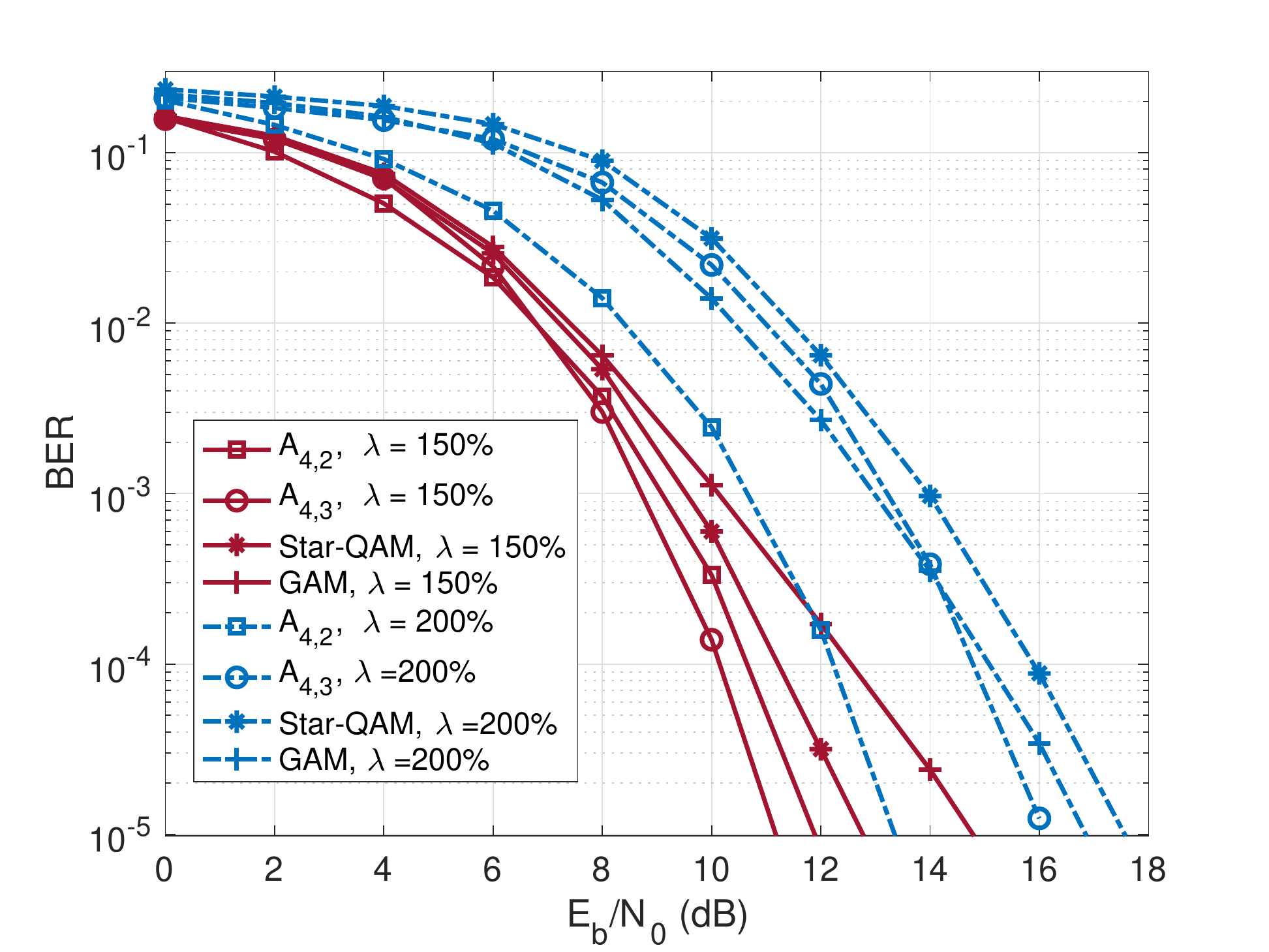}  } 
\subcaptionbox{$\lambda =150\%$ and $M \in \left\{8, 16 \right\}$ \label{fig:Ex_Im2}}{\includegraphics[width=0.42\textwidth]{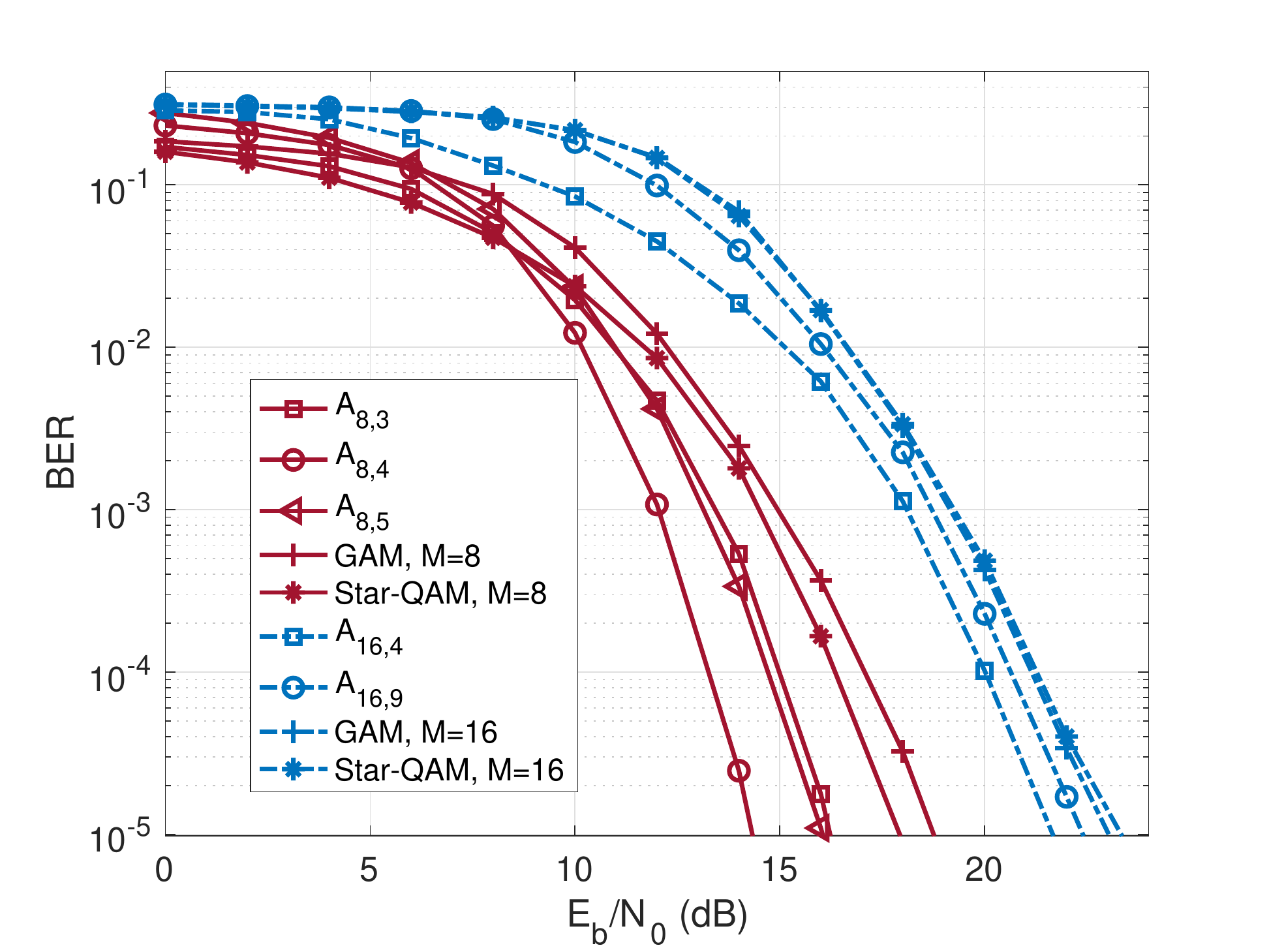}}
\caption{BER performance of different codebooks   with $\kappa \rightarrow \infty$.}
\label{BER_AWGN}
\end{figure*}
 \begin{figure*}
\centering
\subcaptionbox{$M=4$, $\lambda \in \left\{150\%, 200\%  \right\}$   \label{fig:Ex_Im}}{\includegraphics[width=0.42\textwidth]{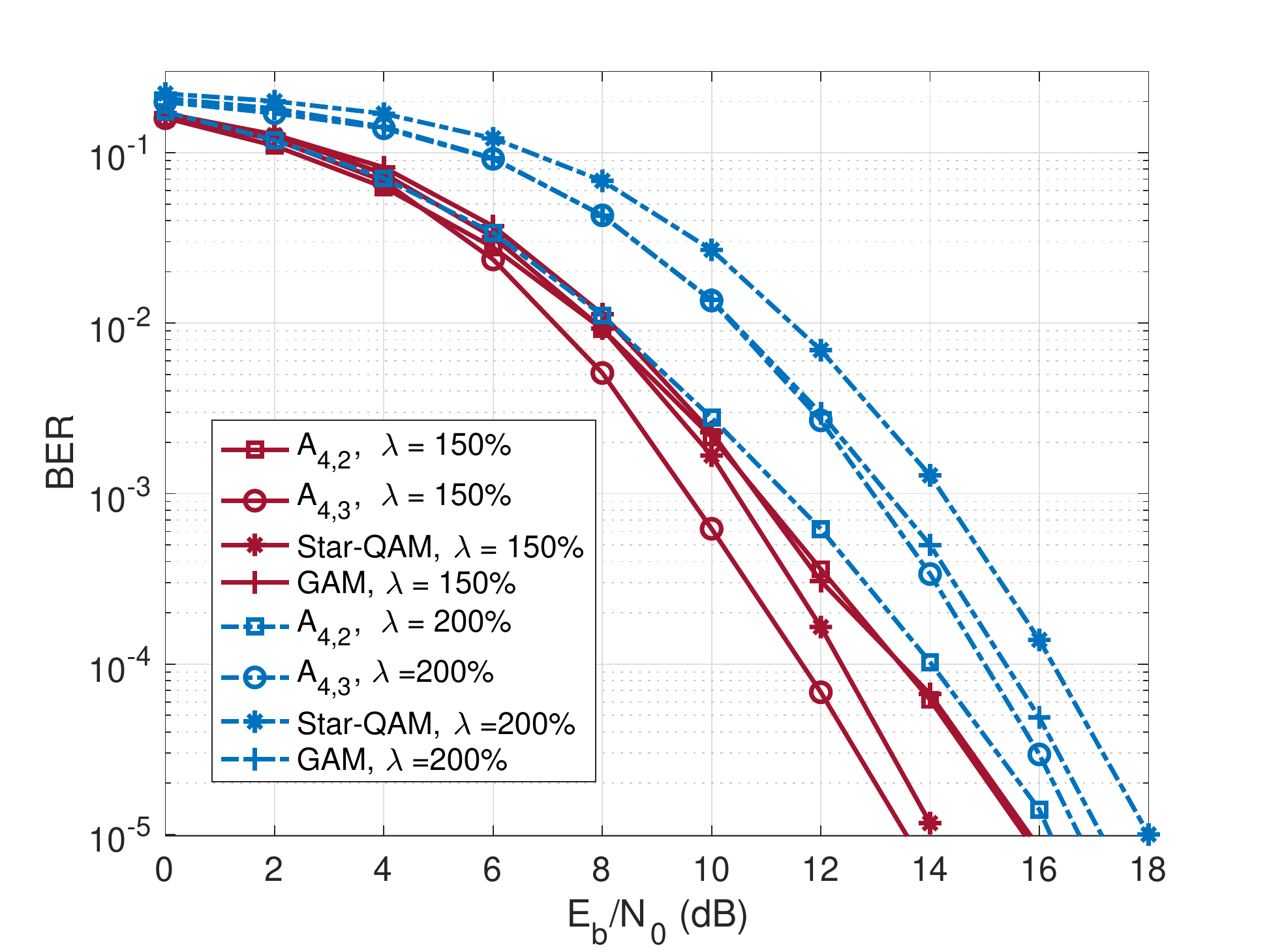}   } 
\subcaptionbox{$\lambda =150\%$ and $M \in \left\{8, 16 \right\}$ \label{fig:Ex_Im2}}{\includegraphics[width=0.42\textwidth]{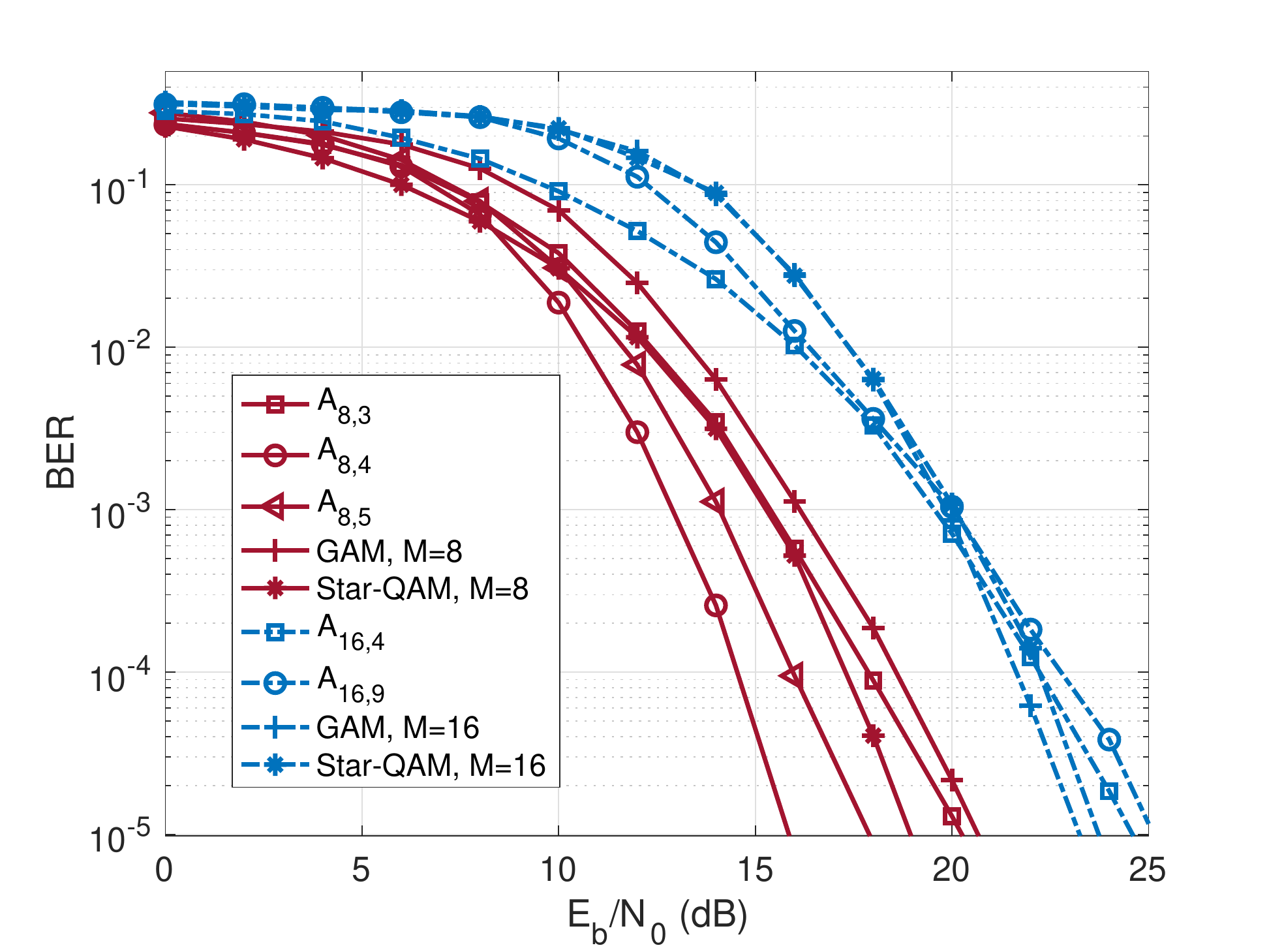} }
\caption{BER performance of different codebooks     with $\kappa= 15$. }\label{BER_rician}
	\vspace{-1.5em}	
\end{figure*}

We first present  the convergence behaviors of different codebooks with $\kappa \rightarrow \infty$ of MPA decoder in Fig. \ref{MPA_iter}.    The number of iterations $I_t$ for decoding convergence of different codebooks are also summarized in Table \ref{codebooks}.
It is noted that most of the proposed codebooks enjoy   faster convergence rate  than the Star-QAM codebook and GAM codebook\footnote{For other values of $\kappa$, we also observe the same result that the proposed LPCBs require less iterations for convergence.}. With reduced  projected numbers ($T$), less message  propagation  between function nodes and variable nodes are required  during MPA iteration, thus leading to less iterations for convergence.  
In particular,   the proposed codebooks $\text{A}_{4,2}$,  $\text{A}_{8,3}$ and  $\text{A}_{16,4}$ for $\lambda = 150\%$,  and $\text{A}_{4,2}$ for $\lambda = 200\%$ only need about one iteration to converge. The decoding can be carried out in a one-shot for these codebooks and the decoding complexity is significantly reduced compared to other codebooks.   

 The computation complexity  in terms of multiplication  and addition of   MPA  can be approximated as \cite{yang2016low,bayesteh2015low}:
\begin{equation} 
\small
\begin{aligned}
&{N_{m}} = [({d_{f}} + 3){T^{d_{f}}}{N} \! + \! ({N} \!- \!2)T{N}]J{I_{t}} + T({N} - 1)J,\\ 
&{N_{a}} 
=[({d_{f}}\! + \!1){T^{d_{f}}}{N} \! +\!  (T \!- \!1){N}{\mathrm{ + (}}{T^{d_f \!- \! 1}} \!-\! 1)T{N}]J{I_{t}},  \\  \end{aligned}
\end{equation}
respectively. Clearly,  the decoding complexity depends on  $T$, $I_{t}$ and the system overloading.     To proceed, we further define the   complexity reduction ratio (CRR) as follows \cite{bao2017bit}:
 \begin{equation}
 \small
     \text{CRR} \triangleq 1- \frac{\text{Number of Operations of LPCBs}}{\text{Number of Operations of Non-LPCBs}}.
 \end{equation}
  
As shown in Fig. \ref{complexity}, it is evident that the proposed LPCBs can greatly reduce the detection complexity, with  more than about $60\%$ complexity reduction being achieved compared to Non-LPCBs.  
The gain becomes more significant  as the increase of $M$ and  $\lambda$.   For example, the proposed codebooks  $\text{A}_{8, 3}$ and $\text{A}_{16,4}$ can reduce the decoding complexity by $97.4\%$ and $99.8\%$, respectively.

 \begin{figure*}[htbp]
\centering
\subcaptionbox{$M=4$ and $\lambda =150\%  $ \label{fig:Ex_Im}}{\includegraphics[width=0.44\textwidth]{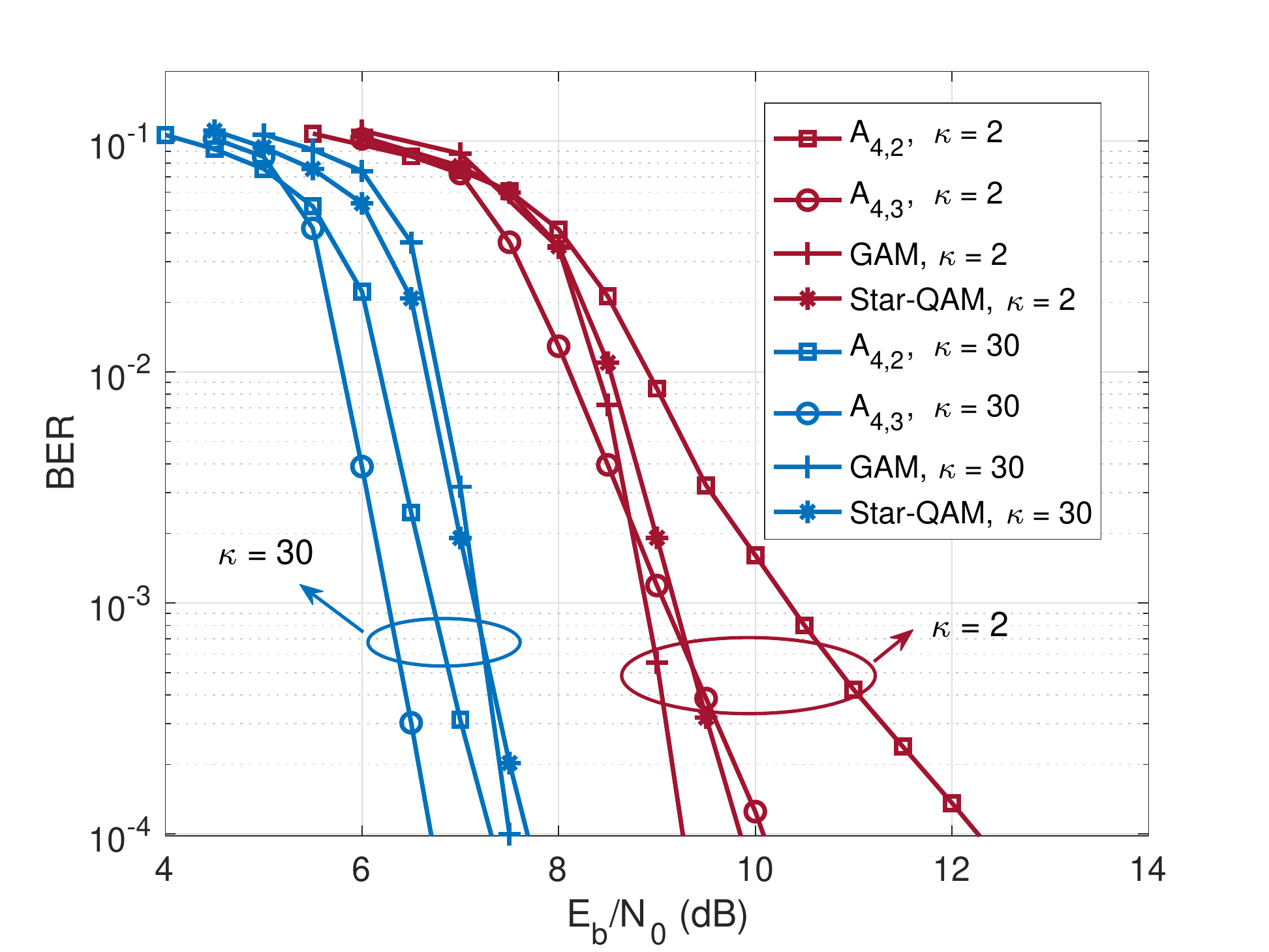}     } 
\subcaptionbox{$M=8$ and $\lambda =150\%  $ \label{fig:Ex_Im2}}{\includegraphics[width=0.44\textwidth]{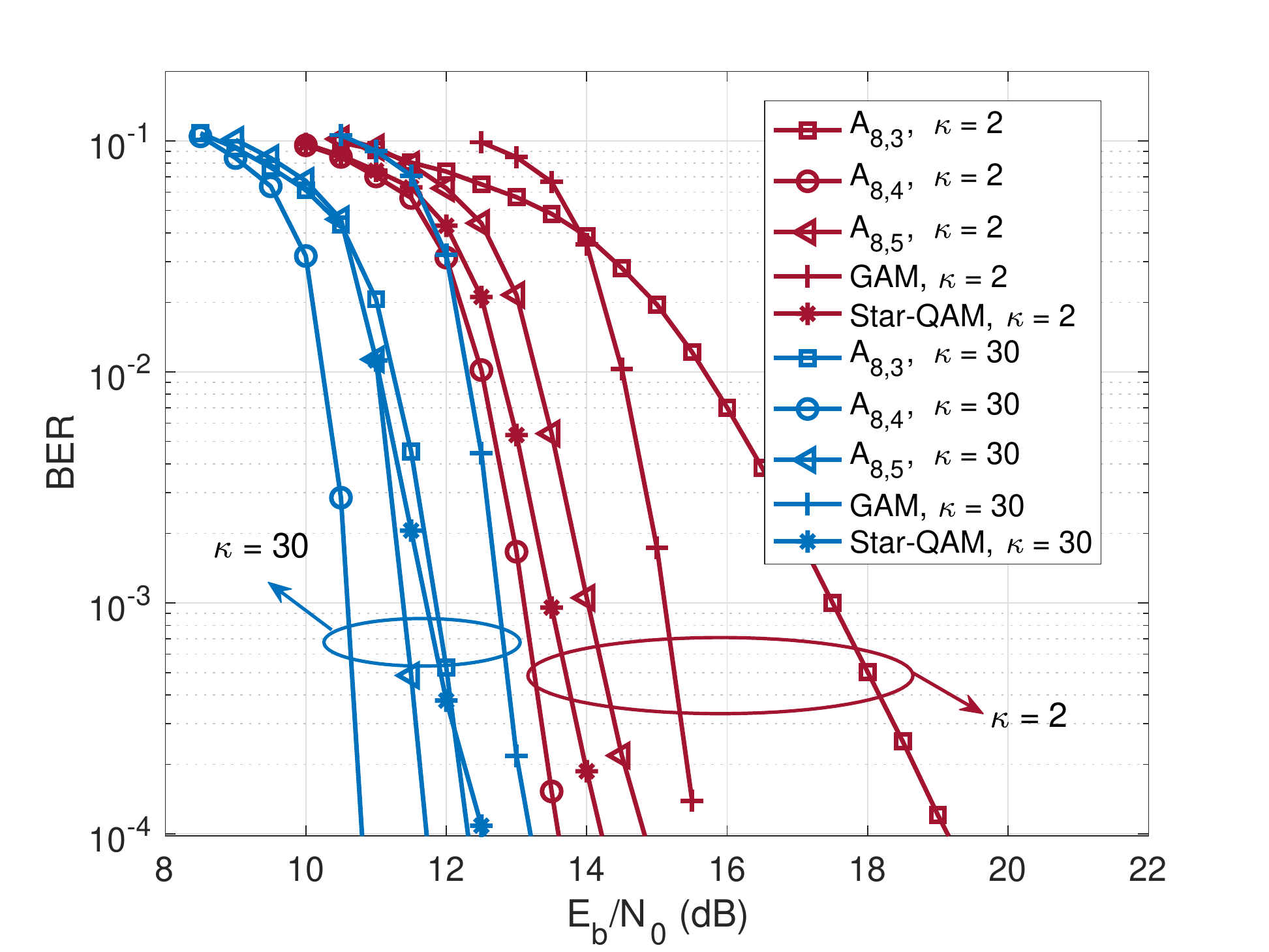}   }
\subcaptionbox{$M=16$ and $\lambda =150\%  $ \label{fig:Ex_Im}}{\includegraphics[width=0.44\textwidth]{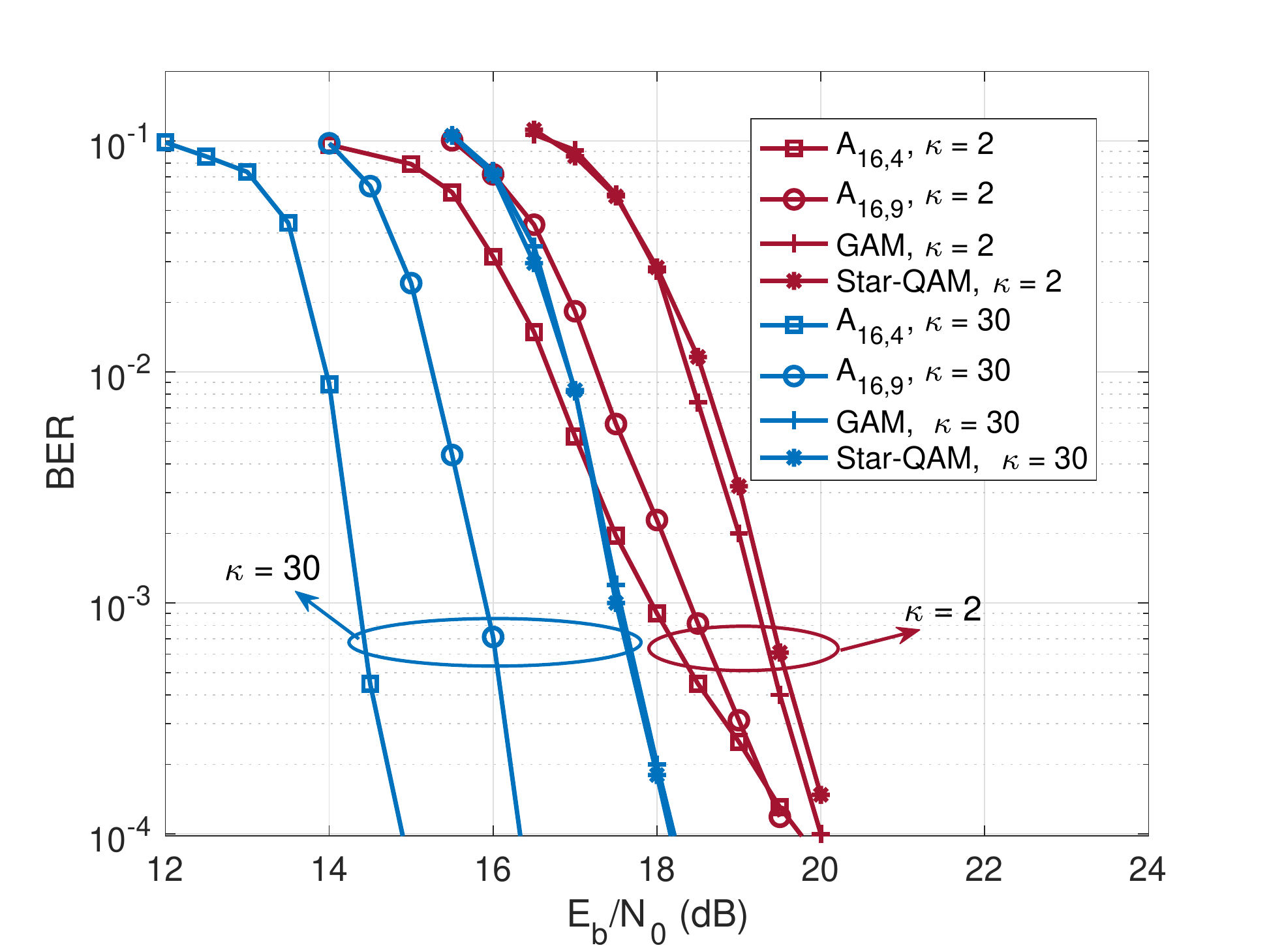}      } 
\subcaptionbox{$M=4$ and $\lambda =200\%  $ \label{fig:Ex_Im2}}{\includegraphics[width=0.44\textwidth]{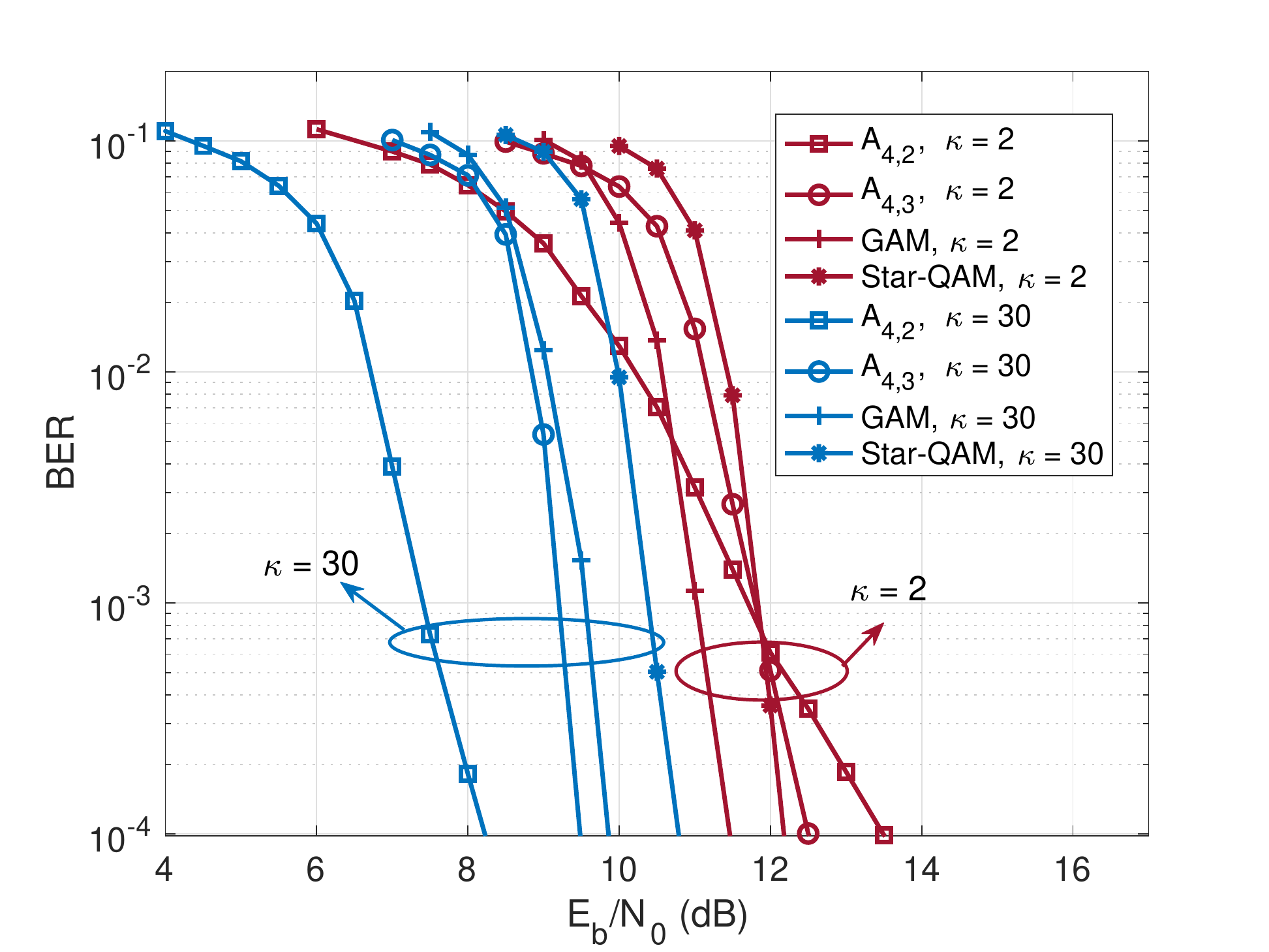}   }
\caption{LDPC coded BER performance comparison of different  codebooks.}\label{CodedBER}
\end{figure*}

\subsection{Comparison of uncoded BER performance}

In this subsection, we evaluate the uncoded BER performance of the proposed  LPCBs.  
    $\kappa \rightarrow \infty$ is the typical value for open environments in satellite IoT \cite{5G_NR_s_iot}. In the rural and suburban scenarios of satellite IoT, and UAV IoT networks,   large  $\kappa$  is  generally assumed \cite{lutz1991land, vucetic1992channel,ErnestNOMA}.    Hence, we consider  $\kappa \rightarrow \infty$ and $\kappa  = 15$ for the uncoded systems, which are shown in Fig. \ref{BER_AWGN} and  Fig. \ref{BER_rician}, respectively.  
   As can be seen in Fig. \ref{BER_AWGN}, the proposed codebook  $\text{A}_{4,3}$   for $\lambda = 150\%$ achieves about $2$ dB gain over the Star-QAM codebook and $4$ dB gain over the GAM codebook at BER = $10^{-5}$, respectively. Moreover,   $\text{A}_{4,3}$   for $\lambda =200\%$ also outperforms  than  the Star-QAM codebook and the GAM codebook. The excellent BER advantage  can be observed for the proposed   $\text{A}_{4,2}$  codebook for $\lambda = 200\%$ due to large MED, which achieves about $3$ dB gain over the Star-QAM codebook and $4.5$ dB gain over the GAM codebook at BER=$10^{-5}$, respectively. For $M \in \left\{8, 16 \right\}$ and $\lambda =150\%$, we are interested in   codebooks with less projected numbers, thus only $\text{A}_{8,3}$, $\text{A}_{8,4}$ and $\text{A}_{8,5}$ are compared to GAM and Star-QAM codebooks in Fig. \ref{BER_AWGN}(b). One can see that  the proposed codebooks outperform the GAM and Star-QAM codebooks significantly. Among these codebooks, $\text{A}_{8,4}$ and $\text{A}_{16,4}$ achieve the best BER performance for $M=8$ and $M=16$, respectively.  In particular,  $\text{A}_{8,4}$  has a gain about $3.8$ dB over the GAM codebook and $4.2$ dB over  the Star-QAM codebook, respectively.

From Fig. \ref{BER_rician},   we have the following main observations: 1) For the case of $\kappa  = 15$, most of the proposed  LPCBs   achieve good BER performance. In particular, the codebooks $\text{A}_{4,3}$ ($\lambda =150\%$), $\text{A}_{4,2}$ ($\lambda =200\%$), $\text{A}_{8,4}$ still outperform the GAM and Star-QAM codebooks;  2) Compared with the results in  Fig. \ref{BER_AWGN}, the performance of  LPCBs with smallest LP numbers for  $\kappa  = 15$, such as $\text{A}_{4,2}$ ($\lambda =150 \%$),  $\text{A}_{8,3}$, $\text{A}_{16,4}$  are more likely affected  by the channel fading.

\subsection{Comparison of coded BER performance}

 \begin{table}  
\small
    \caption{Simulation Parameters}
    \centering
    \begin{tabular}{c|c}
    \hline
     \hline
       \textbf{ Parameters}  &  \textbf{ Values}  \\
        \hline
         \hline
       Transmission  &  Downlink \\
        \hline
       Channel model  &  Rician fading channel, $\kappa =2, 30$ \\
        \hline
        SCMA setting & $\lambda = 150\%$ and $\lambda = 200\%$ \\
          \hline
        Channel coding &  \makecell[c] {5G NR LDPC codes with \\ rate  of $0.8462$ and block length of $260$ }  \\ 
         \hline
        Codebooks &  \makecell[c] {The same codebooks presented in  \\ Fig. \ref{BER_AWGN} and Fig. \ref{BER_rician} }  \\ 
       \hline
       Receiver &   \makecell[c] { Turbo-MPA: \\ $1$ MPA iteration  for  $\text{A}_{4,2}$ ($\lambda = 150\%$), \\ $\text{A}_{4,2}$ ($\lambda = 200\%$), $\text{A}_{8,3}$ and $\text{A}_{16,4}$, \\and $4$ MPA iterations for other codebooks, \\ 5 BICM iterations } \\
         \hline
    \end{tabular}
    \label{sim_para}
    \vspace{-1.5em}
\end{table}

In this subsection, we compare the coded BER performances of the    proposed codebooks with the GAM   and Star-QAM codebooks.  In some scenarios, such as   dense urban and urban areas with an LoS path,      $\kappa$ is relatively small \cite{5G_NR_s_iot, lutz1991land, vucetic1992channel}. Hence, we set  $\kappa$ with a wide range. Considering  the short-packet nature of IoT networks, we apply the 5G NR LDPC codes with small block length \cite{liu2021sparse,PerturbedDeng}, as specified in TS38.212 \cite{5G_NR}.  The detailed simulation settings are  given in Table \ref{sim_para} and the simulated   BER performance   are depicted in Fig. \ref{CodedBER}.

In Fig. \ref{CodedBER}(a), the proposed codebook $\text{A}_{4,2}$ and  $\text{A}_{4,3}$  show  better performance than the GAM and Star-QAM codebooks for $\kappa = 30$. In particular, the  $\text{A}_{4,3}$ outperforms Star-QAM codebook by $1$ dB gain at BER$=10^{-4}$. When there exists severe fading, i.e., $\kappa =2$, the error performance  of $\text{A}_{4,2}$ deteriorates, however, the  proposed $\text{A}_{4,3}$ still achieves similar BER performance with the Star-QAM codebook and performs  slightly better in the low $E_b/N_0$ range. By observing the slopes   of the curves, the GAM codebook enjoys a larger   diversity gain than the low projected codebooks.  This is because some constellation points in the LPCBs are overlapped in order to reduce the decoding complexity, thus the diversity between these constellation pairs may be reduced.   However, we show that by properly designing the MC and maximising the coding gain, most of the proposed LPCBs  can still achieve good   BER performance for $\kappa =2$.

For the    $8$-ary and $16$-ary  codebooks shown in Fig.  \ref{CodedBER}(b) and Fig.  \ref{CodedBER}(c), respectively, BER trends similar to that of the $4$-ary codebook are observed.   The proposed   codebooks outperform  the GAM  and Star-QAM codebooks for  $\kappa=30$, whereas the error  performance of the  proposed codebooks with   least projection numbers  deteriorate  for the case of $\kappa =2$.  
It is interesting to  see the proposed codebook $\text{A}_{8,4}$ and $\text{A}_{16,4}$ achieve the best coded BER performance for both  $\kappa=30$ and  $\kappa=2$ among all the codebooks owing to the well optimized bit labeling   and large coding gain. For example, $\text{A}_{8,4}$ achieves about $3.6$ dB gain over the GAM codebook and $1.5$ dB gain over the Star-QAM  codebook for $\kappa=30$,  and   achieves about $2$ dB gain over the GAM codebook and $1$ dB gain over the Star-QAM  codebook for $\kappa=2$ at BER$=10^{-4}$, respectively.
 
 As shown in Fig.  \ref{CodedBER}(d), for the  SCMA system with $\lambda = 200\%$,   the proposed codebook $\text{A}_{4,2}$ achieves about $1$ dB gain compared to the GAM with $\kappa = 30$ due to the very large coding gain, however, the BER curve degrades   earlier than others for    $\kappa=2$. The  superiority of  the proposed codebooks mainly   lies in  the huge reduction of complexity, especially in largely  overloaded system.

\section{Conclusion}
 
In this paper,  we have  introduced a novel  class of SCMA codebooks that  can significantly reduce the decoding complexity  while achieving  good BER performance
  for downlink IoT networks.  We    have analyzed  the corresponding pairwise  error probability   of SCMA transmissions under a generalized  Rician fading channel model and derived the codebook design criteria.   To reduce the decoding complexity, we have proposed to construct the MC with LP-GAM by allowing several overlapped constellation points in each dimension.   In addition, we have developed     an  efficient  codebook design approach, including permutation, bit labeling, sparse codebook construction and parameter optimization. 
Numerical results demonstrated the  benefits of proposed codebook in terms of complexity reduction and BER performance improvements in different  IoT  environments. In particular, some codebooks (e.g., $\text{A}_{4,2}$, $\text{A}_{4,3}$, $\text{A}_{8,4}$ and $\text{A}_{16,4}$ for   $ \lambda = 150\%$ system  and $\text{A}_{4,2}$ for $ \lambda = 200\%$ system ) achieve significant performance improvements for typical $\kappa $ values in  open, rural and suburban environments,  and significant complexity reduction compared  to the existing codebooks.


%

 \appendices
\section{Proof of the Lemma 1} 
\label{AppeA}
\textit{1). }   For  $(1+\kappa) \gg \frac{ \tau _{{{\mathbf w}} \rightarrow {\tilde{\mathbf{w}}}}(k)  }{4N_0}$, we have 
  \begin{equation}
   \small
  \label{d_awgn}
d_{1, \mathbf{w} \to \mathbf{\hat{w}} }^2(k) = \frac{\kappa \tau _{{{\mathbf w}} \rightarrow {\tilde{\mathbf{w}}}}(k) }{\kappa +1}, d_{2, \mathbf{w} \to \mathbf{\hat{w}} }^2(k)=0,
    \end{equation}
and thus $d_{ \mathbf{w} \to \mathbf{\hat{w}} }^2 = \frac{\kappa \delta_{{{\mathbf w}} \rightarrow {\tilde{\mathbf{w}}}}}{1+\kappa}$. In this special case, (\ref{PEP_Raician2})   reduces to 
\begin{equation}
 \small
 \label{PEP_awgn}
\begin{aligned} 
\text{Pr} \{\mathbf{w} \to \mathbf{\hat{w}}\}   \leq  \exp \left(-\frac{\kappa \delta_{{{\mathbf w}} \rightarrow {\tilde{\mathbf{w}}}}}{4\left( 1+\kappa \right)N_0}\right).
\end{aligned}
  \end{equation}
  
Then, based on  (\ref{ABER}) and  (\ref{PEP_awgn}) , we know that the ABER is dominated by the  MED  between $ \mathbf {w} $ and $\hat {\mathbf {w}}$ in $\Phi _{M^J}$, which corresponding to the MEEs that  all the errors occurred with multiple users.  Therefore,   an important codebook design criteria in this case   is  to  maximize  the MED of the superimposed codewords, which has been widely employed  for SCMA systems operating in the   AWGN channel \cite{klimentyev2017scma,huang2021downlink,li2020design}.  Note that the MED can also obtained by letting   $\kappa \rightarrow \infty$ in $ {{\Delta}_{\min} \left( { \boldsymbol {\mathcal X}} \right)}$.

\textit{2). } In the case when   $(1+\kappa) \ll \frac{ \tau _{{{\mathbf w}} \rightarrow {\tilde{\mathbf{w}}}}(k)  }{4N_0}$,  we have
  \begin{equation}
   \small
  \label{dRay}
d_{1, \mathbf{w} \to \mathbf{\hat{w}} }^2(k) =4N_0 \kappa, d_{2, \mathbf{w} \to \mathbf{\hat{w}} }^2(k)= 4N_0   \ln \left(- \frac{{  \tau _{{{\mathbf w}} \rightarrow {\tilde{\mathbf{w}}}}(k)  }  }{4N_0 \left( 1+ \kappa\right)}   \right),
    \end{equation}
and thus $d_{ \mathbf{w} \to \mathbf{\hat{w}} }^2$ is the sum of the logarithms of the element-wise distance. 
Substituting (\ref{dRay}) and (\ref{d}) into (\ref{PEP_Raician2}), we obtain the PEP   for this special   case as 
    \begin{equation}
\label{PEP_rayleigh}
 \small
\begin{aligned} 
 \small
\text{Pr} \{\mathbf{w}\! \to \!\mathbf{\hat{w}}\}   \leq 
 {\prod \limits _{k \in D (\mathbf {w}\rightarrow \hat {\mathbf {w}})}  \! \left( \frac{   \tau _{{{\mathbf w}} \rightarrow {\tilde{\mathbf{w}}}}(k)   }{4N_0\left(1+\kappa\right)}  \right)^{-1} \!
 {\exp \left(- \kappa \right) }}. 
\end{aligned}
  \end{equation}
where $ D (\mathbf {w}\rightarrow \hat {\mathbf {w}})$ denotes the     set of  indices in which $ \tau _{{{\mathbf w}} \rightarrow {\tilde{\mathbf{w}}}}(k)   \neq 0$.
Note  that one can approximate the right-hand side of (\ref{PEP_rayleigh})  as  
 \begin{equation}
  \small
 \label{pep_gcgd}
\text{Pr} \{\mathbf{w} \!\to \!\mathbf{\hat{w}}\} \leq
G_{c} \left(\mathbf {w} \! \rightarrow \!\hat{\mathbf {w}} \right) \!
  \left (\frac {1}{ 4N_0 \left(1\!+\! \kappa\right)} \right)^{-{G_{d} (\mathbf {w}\rightarrow \hat {\mathbf {w}})}}, 
 \end{equation}
where ${G_{d} (\mathbf {w}\rightarrow \hat {\mathbf {w}})}$ is the cardinality of set $D(\mathbf {w}\rightarrow \hat {\mathbf {w}})$ and
\begin{equation}
 \small
\label{PEP_rayleigh_coding}
\begin{aligned} 
G_{c} \left(\mathbf {w}\rightarrow \hat{\mathbf {w}} \right)=
 {\prod \limits _{k \in D (\mathbf {w}\rightarrow \hat {\mathbf {w}})}      \tau _{{{\mathbf w}} \rightarrow {\tilde{\mathbf{w}}}}(k)    ^{-1}
 {\exp \left(- \kappa \right) }}. 
\end{aligned}
  \end{equation}
   
   Let $G_{d}\triangleq \min _{\mathbf {w}\neq \hat {\mathbf {w}}}G_{d} (\mathbf {w}\rightarrow \hat {\mathbf {w}})$ and $G_{c}\triangleq \min _{\mathbf {w}\neq \hat {\mathbf {w}}}G_{c} (\mathbf {w}\rightarrow \hat {\mathbf {w}})$, where $G_{d}$ and $G_{c}$ are the so-called   diversity gain and coding gain \cite{xin2003space}.  
     For  $\kappa \rightarrow 0$,      from (\ref{pep_gcgd})   we observe  the PEP  decays with the slope of $\left ( 1/{\left( \kappa +1\right)N_0} \right)^{-{G_{d}}}$ and the system achieves the diversity order of $G_{d}$. The ABER is mainly dominated by the SEE that all the codewords are detected correctly except for  a codeword of only one user. In other words,  for the pairs $ \mathbf {w} $ and $\hat {\mathbf {w}} $ such that $ G_{d} (\mathbf {w}\rightarrow \hat {\mathbf {w}}) = G_{d}$,    the  term ${\prod \nolimits _{k \in D_k (\mathbf {w}\rightarrow \hat {\mathbf {w}})}      \tau _{{{\mathbf w}} \rightarrow {\tilde{\mathbf{w}}}}(k)    ^{-1}}$ in  (\ref{PEP_rayleigh_coding}) equals to the MPD of the sparse codebook. The MPD of the codebook  is given by 
     \begin{equation} 
 \small
\text{MPD}(\boldsymbol {\mathcal X})= \underset {1\leq j\leq J} {\min }     \underset {i\neq l}{\min } \prod _{k \in \rho(\mathbf {x}_i^{(j)}, {\mathbf {x}}_l^{(j)})} |x_{k,i}^{(j)}-x_{k,l}^{(j)}|^{2},
\end{equation}
  where $x_{k,i}^{(j)}$ is the $j$th user's $i$th codeword at $k$th entry and
  $\rho (\mathbf {x}_i^{(j)}, {\mathbf {x}}_l^{(j)})$ denotes the set of  indices in which $ {x}_{k,i}^{(j)} \neq {{x}}_{k,j}^{(j)}$. 
 Similarly, in  the case when $ G_{d} (\mathbf {w}\rightarrow \hat {\mathbf {w}}) = G_{d}$ and let $\kappa =  0$ (Rayleigh fading channel),  ${{\Delta}_{\min} \left( { \boldsymbol {\mathcal X}} \right)}$ reduces to  
 \begin{equation} 
 \small
 \label{k0}
 \begin{aligned}
{{\Delta}_{\min}^{\kappa = 0} \left( { \boldsymbol {\mathcal X}} \right)} &  =    \underset {1\leq j\leq J} {\min }    \underset {i\neq l}{\min }  \quad \sum \limits _{k=1}^K   { 4N_0   \ln \left(1+  \frac{{  |x_{k,i}-x_{k,l}|^{2} }  } {4N_0}   \right) }\\
& \approx 4N_0 \ln \left(\text{MPD}(\boldsymbol {\mathcal X})  (4N_0)^{-G_d}  \right).
 \end{aligned}
  \end{equation}

For $\kappa=0$,  (\ref{k0})   indicates that to improve the ABER, it is idea to improve the $G_d$ first and then maximize the $\text{MPD}(\boldsymbol {\mathcal X})$ \cite{xin2003space}.

  This completes the proof of Lemma 1.

\section{Proof of the Lemma 2}
\label{AppeB}
   For an $M$-ary  $  \boldsymbol {\mathcal C}_{MC}$,  the primary design rule is that the resultant MC should be uniquely separable.  Namely, giving arbitrary two vectors $\mathbf {\bar c}_{i}  $ and $ \mathbf {\bar c}_{l} $  in  $  \boldsymbol {\mathcal C}_{MC}$, we have  $\mathbf {\bar c}_{i}  \neq \mathbf {\bar c}_{l} $ with $ i \neq l$. Hence, for the basic constellation with $T$ distinct points, the $N$-dimensional constellation with maximum size can be constructed is $T^N$, where construction is given by the Cartesian product over the basic  constellation. For example,  consider $T=2$ and $N=2$  and assume that the designed basic constellation $\mathcal A_2 = \{-a, +a\}  \in \mathbf {\mathbb C}$, obviously, the unique decodable $2$-dimensional constellation with maximum size  can be  generated as 
   $ \mathcal A_2 \times \mathcal A_2  \rightarrow \boldsymbol {\mathcal C}_{MC} $, i.e.,
   \begin{equation}
\begin{aligned} \boldsymbol {\mathcal C}_{MC}=\left [{ \begin{matrix} -a&+a&-a&+a\\ -a&-a&+a&+a \end{matrix} }\right].
\end{aligned}
\end{equation}
 
Therefore, for the given $M$ and $N$, the LP number should satisfy $\left\lceil   {\sqrt[N]{M}}\right\rceil  \leq T \leq M$, and for the case  of $ {T = \sqrt[N]{M}} \in \mathbb Z$, the construction of $ \boldsymbol {\mathcal C}_{MC} $ is given by the Cartesian product of $\mathcal A _{{T}}$.
   
This completes the proof of Lemma 2.

\section{Proof of the Lemma 4} 
\label{AppeC}

The proof of Lemma 4 can be proceed in two steps as follows.

 In the first step, we show that $\boldsymbol {\Phi}_{M^J}$ can be expressed as the Cartesian product  of $\boldsymbol{\mathcal S}_{\text{sum}}$. To proceed,  we  rewrite the constellation superposition set as 
 \begin{equation}
 \small
    \Phi_{M^J}  =  \left\{ \mathbf w = \mathbf x_1 + \mathbf x_2 + \cdots  +\mathbf x_J \vert \forall \mathbf x_j\in  \boldsymbol {\mathcal{X}}_{j}, j =1,2,\ldots,J\right\}, 
 \end{equation}
 where the mapping rule is given by $ g_{\Phi}: \boldsymbol{\mathcal X}_{1}+  \cdots +  \boldsymbol{\mathcal X}_{J} \rightarrow \Phi_{M^J} $.   In fact,  the constellation  superposition process is characterized by a Cartesian product mapping. Similarly, we have  $ g_{\boldsymbol{\mathcal S}}: z_1  \mathcal A _{{T}} + \cdots + z_{d_f} \mathcal A _{{T}} \rightarrow \boldsymbol{\mathcal S}_{\text{sum}}$.  
 The lemma $2$ reveals that for $  {T = \sqrt[N]{M}}$, the mapping  $ f_{\text{MC}}: \mathcal A _{{T}} \times \cdots \times \mathcal A _{{T}} \rightarrow \boldsymbol {\mathcal C}_{\text{MC}}$ is  a Cartesian product. Accordingly,  the mapping $ f_{\boldsymbol{\mathcal{C}}_j }:  z_1^{(j)}  \mathcal A _{{T}} \times \cdots \times z_{N}^{(j)} \mathcal A _{{T}}  \rightarrow \boldsymbol {\mathcal C}_{j}$ is also a Cartesian product, where $z_{n}^{(j)}$ is the    $j$th user's constellation operator. Based on $f_{\mathcal{C}_j }$, we can rewrite $g_{\Phi}$ as
 \begin{equation}
 \small\label{gf}
    g_{\Phi}: \mathbf{V}_{1}f_{\boldsymbol{\mathcal{C}}_j }  +  \cdots +   \mathbf{V}_{J}f_{\boldsymbol{\mathcal{C}}_j } \rightarrow \Phi_{M^J}.   
 \end{equation}
 
  (\ref{gf}) indicates that  $\Phi_{M^J}$  is obtained by the two  steps:  1) Mapping   the $1$-dimension LP-GAM $\mathcal A _{{T}}$ to  $ \boldsymbol {\mathcal C}_{j}$; 2) Constellation superposition. Since the two steps are both characterized by the Cartesian product, the changes in the order of the two steps will not affect  the result of $ \Phi_{M^J}$. Namely,    $ \Phi_{M^J}$ can also be obtained by: 1)  Constellation superposition   based on LP-GAM, i.e., $ g_{\boldsymbol{\mathcal S}}: z_1  \mathcal A _{{T}} + \cdots + z_{d_f} \mathcal A _{{T}} \rightarrow \boldsymbol{\mathcal S}_{\text{sum}}$; 2) Obtain  $ \Phi_{M^J}$ by performing  Cartesian product over $\boldsymbol{\mathcal S}_{\text{sum}}$, i.e,  
  \begin{equation}
 \small\label{gf2}
    f_{\Phi}: \boldsymbol{\mathcal S}_{\text{sum}}^{(1)} \times  \cdots \times \boldsymbol{\mathcal S}_{\text{sum}}^{(K)}  \rightarrow \Phi_{M^J}.   
 \end{equation}
 
 In Step 2,  we show that maximize   $\text{MED} \big ({\boldsymbol{\mathcal S}_{\text{sum}} } \big) $ is equivalent to maximizing    $\Delta_{\min}\left( {   \boldsymbol {\mathcal X}_{\lambda,M,T}      }\right) $.  Let ${{\mathbf{w}} }$ and ${ \hat{\mathbf{w}} }$    be   two superimposed codeword vectors, and define   \begin{equation}
    \small
    \label{gg}
  G    \triangleq \underset{ {{\mathbf{w}} },  {\hat {\mathbf{w}} },   {{\mathbf{w}} } \neq  {\hat {\mathbf{w}} }
  }{\min}  \underbrace{\sum \limits _{k=1}^{K}\text {Ind}\left (\tau _{{{\mathbf{w}} } \rightarrow {\hat {\mathbf{w}} }}(k)\right)  }_{G({{{\mathbf{w}} } \rightarrow {\hat {\mathbf{w}} }})}, 
  \end{equation} 
  where $\tau _{{{\mathbf{w}} } \rightarrow { \hat{\mathbf{w}} }}(k) = \vert w(k) - \hat w(k) \vert^2 $, and $w(k)$ is  the $k$th entry of $\mathbf w$.  Obviously, since $\Phi_{M^J}$ can be  generated by the Cartesian product of $\boldsymbol{\mathcal S}_{\text{sum}}$,  we have  $ G =1$ and the MED of $\Phi_{M^J}$   equals to $\text{MED} \big ({\boldsymbol{\mathcal S}_{\text{sum}} } \big) $. Assume ${{\mathbf{w}}_m }$ and ${ {\mathbf{w}}_n }$ are the two vectors that achieve  the MED value, and $\tau _{{{\mathbf{w}}_m } \rightarrow { {\mathbf{w}_n} }}(k') \neq 0$. Accordingly, we obtain $\tau _{{{\mathbf{w}}_m } \rightarrow { {\mathbf{w}_n} }}(k') = \text{MED} \big ({\boldsymbol{\mathcal S}_{\text{sum}} } \big)$. This also indicates that for  arbitrary two vectors ${{\mathbf{w}} }$ and ${ \hat{\mathbf{w}} }$,    $\tau _{{{\mathbf{w}} } \rightarrow {\hat {\mathbf{w}} }}(k) = 0$ or $\tau _{{{\mathbf{w}} } \rightarrow {\hat {\mathbf{w}} }}(k) \geq \text{MED} \big ({\boldsymbol{\mathcal S}_{\text{sum}} } \big)$, $k =1, 2, \ldots, K$.

  Based on (\ref{d}), $d_{ \mathbf{w} \to \mathbf{\hat{w}} }^2 $ is an increasing function of ${G({{{\mathbf{w}} } \rightarrow {\hat {\mathbf{w}} }})}$ and $\tau _{{{\mathbf{w}}_m } \rightarrow { {\mathbf{w}_n} }}(k)$. Hence,   (\ref{delta1}) achieves the minimum value at  ${{\mathbf{w}}_m }$ and ${ {\mathbf{w}}_n }$, where $G=1$ and $\tau _{{{\mathbf{w}}_m } \rightarrow { {\mathbf{w}_n} }}(k') = \text{MED} \big ({\boldsymbol{\mathcal S}_{\text{sum}} } \big)$. Therefore, improving $\text{MED} \big ({\boldsymbol{\mathcal S}_{\text{sum}} } \big)$ is equivalent to  increasing  $\Delta_{\min}\left( {   \boldsymbol {\mathcal X}_{\lambda,M,T}      }\right) $.
  
  This completes the proof of Lemma 4.
  \label{AppeD}

  \section{The proposed  LPCBs} 
  The proposed LPCBs of  $\text{A}_{4,3}$ ($\lambda = 150\%$), $\text{A}_{4,2}$ ($\lambda = 200\%$) and $\text{A}_{8,4}$ ($\lambda = 150\%$) are presented. 
  For  $M=4$, the four columns of in $\mathcal X_j$  denote the codewords with labelled by $00$, $01$, $10$, $11$. Similarly, for $M=8$, the  eight   columns represent the codewords labelled by $000$, $001$, $010$, $011$, $100$, $101$, $110$, $111$. 
  
  \vspace{-0.5em}
  \subsection{The proposed LPCB $\text{A}_{4,3}$ ($\lambda = 150\%$)} 
 \begin{equation}
  \footnotesize   
     \mathcal X_1=
\begin{bmatrix}
     0&  0.0850 + 1.0324i&  0& 0\\
   0&  0&  0&  1.0841 + 0.0000i\\
   0& 0&  0& -1.0841 + 0.0000i\\
   0& -0.0850 - 1.0324i&  0&  0\\
      \end{bmatrix}^{\text{T}}, \notag
  \end{equation}
\begin{equation}
  \footnotesize   
     \mathcal X_2=
\begin{bmatrix}
     0.0850 + 1.0324i&  0& 0&  0\\
   0&  0&  1.0841 + 0.0000i&  0\\
  0&  0& -1.0841 + 0.0000i&  0\\
  -0.0850 - 1.0324i&  0&  0&  0\\
      \end{bmatrix}^{\text{T}}, \notag
  \end{equation}
 \begin{equation}
  \footnotesize   
     \mathcal X_3=
\begin{bmatrix}
    -0.7156 + 0.4894i&  0&  0&  0\\
  0& -0.7156 + 0.4894i&  0&  0\\
   0&  0.7156 - 0.4894i&  0&  0\\
   0.7156 - 0.4894i& 0&  0&  0\\
      \end{bmatrix}^{\text{T}}, \notag
  \end{equation}
\begin{equation}
  \footnotesize   
     \mathcal X_4=
\begin{bmatrix}
     0&  0& -0.7156 + 0.4894i&  0\\
   0&  0& 0& -0.7156 + 0.4894i\\
   0&  0&  0&  0.7156 - 0.4894i\\
   0&  0&  0.7156 - 0.4894i& 0\\
      \end{bmatrix}^{\text{T}}, \notag
  \end{equation}
 \begin{equation}
  \footnotesize   
     \mathcal X_5=
\begin{bmatrix}
     1.0841 + 0.0000i&  0&  0& 0\\
   0&  0&  0&  0.0850 + 1.0324i\\
  0&  0&  0& -0.0850 - 1.0324i\\
  -1.0841 + 0.0000i&  0&  0&  0\\
      \end{bmatrix}^{\text{T}}, \notag
  \end{equation}
\begin{equation}
  \footnotesize   
     \mathcal X_6=
\begin{bmatrix}
     0&  1.0841 + 0.0000i& 0&  0\\
   0&  0&  0.0850 + 1.0324i&  0\\
   0& 0& -0.0850 - 1.0324i&  0\\
   0& -1.0841 + 0.0000i&  0&  0\\
      \end{bmatrix}^{\text{T}}. \notag
  \end{equation}

  \subsection{The proposed LPCB $\text{A}_{4,2}$ ($\lambda = 200\%$)}
    \vspace{-0.5em}


 \begin{equation}
  \footnotesize   
     \mathcal X_1=
\begin{bmatrix}
   0.6576 + 0.6755i&  0.5852 - 0.5696i&  0&  0&  0\\
   0.6576 + 0.6755i& -0.5852 + 0.5696i&  0&  0&  0\\
  -0.6576 - 0.6755i&  0.5852 - 0.5696i&  0&  0&  0\\
  -0.6576 - 0.6755i& -0.5852 + 0.5696i&  0&  0&  0\\
      \end{bmatrix}^{\text{T}}, \notag
  \end{equation}
\begin{equation}
  \footnotesize   
     \mathcal X_2=
\begin{bmatrix}
  -0.1282 + 0.4536i&  0& -0.3288 - 0.3378i&  0&  0\\
  -0.1282 + 0.4536i&  0&  0.3288 + 0.3378i&  0&  0\\
   0.1282 - 0.4536i&  0& -0.3288 - 0.3378i&  0&  0\\
   0.1282 - 0.4536i&  0&  0.3288 + 0.3378i&  0&  0\\

      \end{bmatrix}^{\text{T}}, \notag
  \end{equation}
 \begin{equation}
  \footnotesize   
     \mathcal X_3=
\begin{bmatrix}
   0.3288 + 0.3378i&  0&  0&  0.1282 - 0.4536i&  0\\
   0.3288 + 0.3378i&  0&  0& -0.1282 + 0.4536i&  0\\
  -0.3288 - 0.3378i&  0&  0&  0.1282 - 0.4536i&  0\\
  -0.3288 - 0.3378i&  0&  0& -0.1282 + 0.4536i&  0\\

      \end{bmatrix}^{\text{T}}, \notag
  \end{equation}
\begin{equation}
  \footnotesize   
     \mathcal X_4=
\begin{bmatrix}
  -0.5852 + 0.5696i&  0&  0&  0& -0.6576 - 0.6755i\\
  -0.5852 + 0.5696i&  0&  0&  0&  0.6576 + 0.6755i\\
   0.5852 - 0.5696i&  0&  0&  0& -0.6576 - 0.6755i\\
   0.5852 - 0.5696i&  0&  0&  0&  0.6576 + 0.6755i\\
      \end{bmatrix}^{\text{T}}, \notag
  \end{equation}
 \begin{equation}
  \footnotesize   
     \mathcal X_5=
\begin{bmatrix}
   0&  0.6576 + 0.6755i&  0.5852 - 0.5696i&  0&  0\\
   0&  0.6576 + 0.6755i& -0.5852 + 0.5696i&  0&  0\\
   0& -0.6576 - 0.6755i&  0.5852 - 0.5696i&  0&  0\\
   0& -0.6576 - 0.6755i& -0.5852 + 0.5696i&  0&  0\\
      \end{bmatrix}^{\text{T}}, \notag
  \end{equation}
\begin{equation}
  \footnotesize   
     \mathcal X_6=
\begin{bmatrix}
   0& -0.1282 + 0.4536i&  0& -0.3288 - 0.3378i&  0\\
   0& -0.1282 + 0.4536i&  0&  0.3288 + 0.3378i&  0\\
   0&  0.1282 - 0.4536i&  0& -0.3288 - 0.3378i&  0\\
   0&  0.1282 - 0.4536i&  0&  0.3288 + 0.3378i&  0\\
      \end{bmatrix}^{\text{T}}, \notag
  \end{equation}
\begin{equation}
  \footnotesize   
     \mathcal X_7=
\begin{bmatrix}
   0&  0.3288 + 0.3378i&  0&  0&  0.1282 - 0.4536i\\
   0&  0.3288 + 0.3378i&  0&  0& -0.1282 + 0.4536i\\
   0& -0.3288 - 0.3378i&  0&  0&  0.1282 - 0.4536i\\
   0& -0.3288 - 0.3378i&  0&  0& -0.1282 + 0.4536i\\
      \end{bmatrix}^{\text{T}}, \notag
  \end{equation}
\begin{equation}
  \footnotesize   
     \mathcal X_8=
\begin{bmatrix}
   0&  0&  0.6576 + 0.6755i&  0.5852 - 0.5696i&  0\\
   0&  0&  0.6576 + 0.6755i& -0.5852 + 0.5696i&  0\\
   0&  0& -0.6576 - 0.6755i&  0.5852 - 0.5696i&  0\\
   0&  0& -0.6576 - 0.6755i& -0.5852 + 0.5696i&  0\\
      \end{bmatrix}^{\text{T}}, \notag
  \end{equation}
\begin{equation}
  \footnotesize   
     \mathcal X_9=
\begin{bmatrix}
   0&  0& -0.1282 + 0.4536i&  0& -0.3288 - 0.3378i\\
   0&  0& -0.1282 + 0.4536i&  0&  0.3288 + 0.3378i\\
   0&  0&  0.1282 - 0.4536i&  0& -0.3288 - 0.3378i\\
   0&  0&  0.1282 - 0.4536i&  0&  0.3288 + 0.3378i\\
      \end{bmatrix}^{\text{T}}, \notag
  \end{equation}
    
\begin{equation}
  \footnotesize   
     \mathcal X_{10}=
\begin{bmatrix}
   0&  0&  0&  0.6576 + 0.6755i&  0.5852 - 0.5696i\\
   0&  0&  0&  0.6576 + 0.6755i& -0.5852 + 0.5696i\\
   0&  0&  0& -0.6576 - 0.6755i&  0.5852 - 0.5696i\\
   0&  0&  0& -0.6576 - 0.6755i& -0.5852 + 0.5696i\\
      \end{bmatrix}^{\text{T}}. \notag
  \end{equation}


  \subsection{The proposed LPCB $\text{A}_{8,4}$ ($\lambda = 150\%$)} 
    \vspace{-0.5em}
 \begin{equation}
  \footnotesize   
     \mathcal X_1=
\begin{bmatrix}
    0    &     0.3202 + 0.3995i    &     0    &     0.0863 - 0.7668i\\
   0    &     0.3202 + 0.3995i    &     0    &    -0.6940 + 0.4385i\\
   0    &    -0.1187 + 0.5316i    &     0    &     0.6940 - 0.4385i\\
   0    &    -0.1187 + 0.5316i    &     0    &    -0.0863 + 0.7668i\\
   0    &     0.1187 - 0.5316i    &     0    &     0.0863 - 0.7668i\\
   0    &     0.1187 - 0.5316i    &     0    &    -0.6940 + 0.4385i\\
   0    &    -0.3202 - 0.3995i    &     0    &     0.6940 - 0.4385i\\
   0    &    -0.3202 - 0.3995i    &     0    &    -0.0863 + 0.7668i
      \end{bmatrix}^{\text{T}}, \notag
  \end{equation}
\begin{equation}
  \footnotesize   
     \mathcal X_2=
\begin{bmatrix}
     0.3202 + 0.3995i    &     0    &     0.0863 - 0.7668i    &     0\\
   0.3202 + 0.3995i    &     0    &    -0.6940 + 0.4385i    &     0\\
  -0.1187 + 0.5316i    &     0    &     0.6940 - 0.4385i    &     0\\
  -0.1187 + 0.5316i    &     0    &    -0.0863 + 0.7668i    &     0\\
   0.1187 - 0.5316i    &     0    &     0.0863 - 0.7668i    &     0\\
   0.1187 - 0.5316i    &     0    &    -0.6940 + 0.4385i    &     0\\
  -0.3202 - 0.3995i    &     0    &     0.6940 - 0.4385i    &     0\\
  -0.3202 - 0.3995i    &     0    &    -0.0863 + 0.7668i    &     0\\
      \end{bmatrix}^{\text{T}}, \notag
  \end{equation}
 \begin{equation}
  \footnotesize   
     \mathcal X_3=
\begin{bmatrix}
    -0.7410 - 0.0249i    &     0.7410 + 0.0249i    &     0    &     0\\
  -0.7410 - 0.0249i    &    -0.4723 - 0.6317i    &     0    &     0\\
  -0.4723 - 0.6317i    &     0.4723 + 0.6317i    &     0    &     0\\
  -0.4723 - 0.6317i    &    -0.7410 - 0.0249i    &     0    &     0\\
   0.4723 + 0.6317i    &     0.7410 + 0.0249i    &     0    &     0\\
   0.4723 + 0.6317i    &    -0.4723 - 0.6317i    &     0    &     0\\
   0.7410 + 0.0249i    &     0.4723 + 0.6317i    &     0    &     0\\
   0.7410 + 0.0249i    &    -0.7410 - 0.0249i    &     0    &     0\\
      \end{bmatrix}^{\text{T}}, \notag
  \end{equation}
\begin{equation}
  \footnotesize   
     \mathcal X_4=
\begin{bmatrix}
  
   0    &     0    &    -0.7410 - 0.0249i    &     0.7410 + 0.0249i\\
   0    &     0    &    -0.7410 - 0.0249i    &    -0.4723 - 0.6317i\\
   0    &     0    &    -0.4723 - 0.6317i    &     0.4723 + 0.6317i\\
   0    &     0    &    -0.4723 - 0.6317i    &    -0.7410 - 0.0249i\\
   0    &     0    &     0.4723 + 0.6317i    &     0.7410 + 0.0249i\\
   0    &     0    &     0.4723 + 0.6317i    &    -0.4723 - 0.6317i\\
   0    &     0    &     0.7410 + 0.0249i    &     0.4723 + 0.6317i\\
   0    &     0    &     0.7410 + 0.0249i    &    -0.7410 - 0.0249i\\
      \end{bmatrix}^{\text{T}}, \notag
  \end{equation}
 \begin{equation}
  \footnotesize   
     \mathcal X_5=
\begin{bmatrix}
    -0.0863 + 0.7668i    &     0    &     0    &    -0.3202 - 0.3995i\\
  -0.0863 + 0.7668i    &     0    &     0    &    -0.1187 + 0.5316i\\
  -0.6940 + 0.4385i    &     0    &     0    &     0.1187 - 0.5316i\\
  -0.6940 + 0.4385i    &     0    &     0    &     0.3202 + 0.3995i\\
   0.6940 - 0.4385i    &     0    &     0    &    -0.3202 - 0.3995i\\
   0.6940 - 0.4385i    &     0    &     0    &    -0.1187 + 0.5316i\\
   0.0863 - 0.7668i    &     0    &     0    &     0.1187 - 0.5316i\\
   0.0863 - 0.7668i    &     0    &     0    &     0.3202 + 0.3995i\\
      \end{bmatrix}^{\text{T}}, \notag
  \end{equation}
\begin{equation}
  \footnotesize   
     \mathcal X_6=
\begin{bmatrix}
   0    &    -0.0863 + 0.7668i    &    -0.3202 - 0.3995i    &     0\\
   0    &    -0.0863 + 0.7668i    &    -0.1187 + 0.5316i    &     0\\
   0    &    -0.6940 + 0.4385i    &     0.1187 - 0.5316i    &     0\\
   0    &    -0.6940 + 0.4385i    &     0.3202 + 0.3995i    &     0\\
   0    &     0.6940 - 0.4385i    &    -0.3202 - 0.3995i    &     0\\
   0    &     0.6940 - 0.4385i    &    -0.1187 + 0.5316i    &     0\\
   0    &     0.0863 - 0.7668i    &     0.1187 - 0.5316i    &     0\\
   0    &     0.0863 - 0.7668i    &     0.3202 + 0.3995i    &     0\\
      \end{bmatrix}^{\text{T}}. \notag
  \end{equation}

\ifCLASSOPTIONcaptionsoff
  \newpage
\fi

 \bibliography{bibtex/bib/ref} 
\bibliographystyle{IEEEtran}








\end{document}